\documentclass{aa}
\usepackage{psfig,graphicx,natbib}
\bibpunct{(}{)}{;}{a}{}{,}
\begin{document}
\title{The multi-phase gaseous halos of star forming late-type galaxies}

\subtitle{I. XMM-Newton observations of the Hot Ionized Medium\thanks{Based on observations obtained with XMM-Newton, an ESA science mission with instruments and contributions directly funded by ESA Member States and NASA.}}

   \author{R. T\"ullmann
          \inst{1}
          \and
          W. Pietsch
          \inst{2}
          \and
          J. Rossa\inst{3}
          \and
          D. Breitschwerdt\inst{4}
          \and
          R.-J. Dettmar\inst{1}
          }

   \offprints{R.~T\"ullmann}

   \institute{Astronomisches Institut, Ruhr-Universit\"at Bochum,
              D-44780 Bochum, Germany\\
              \email{tullmann@astro.rub.de,dettmar@astro.rub.de}  
          \and
              Max-Planck Institut f\"ur extraterrestrische Physik,
              Geissenbachstrasse, 85748 Garching, Germany\\
              \email{wnp@mpe.mpg.de}
          \and
              Space Telescope Science Institute, 3700 San Martin Drive,  
              Baltimore, MD 21218, U.S.A.\\ 
              \email{jrossa@stsci.edu}
          \and 
              Institut f\"ur Astronomie, T\"urkenschanzstrasse 17, A-1180 Wien, Austria\\ 
              \email{breitschwerdt@astro.univie.ac.at}
              }
   \date{Received February 25, 2005; accepted September 25, 2005}

   \abstract{This study presents first results from an X-ray mini-survey carried out with XMM-Newton to investigate the diffuse Hot Ionized Medium in the halos of nine nearby star-forming edge-on spiral galaxies. Diffuse gaseous X-ray halos are detected in eight of our targets, covering a wide range of star formation rates from quiescent to starburst cases. For four edge-on spiral galaxies, namely NGC\,3044, NGC\,3221, NGC\,4634, and NGC\,5775, we present the first published high resolution/sensitivity detections of extended soft X-ray halos. {\small EPIC} X-ray contour maps overlaid onto H$\alpha$ imaging data reveals that in all cases the presence of X-ray halos is correlated with extraplanar Diffuse Ionized Gas. Moreover, these halos are also associated with non-thermal cosmic ray halos, as evidenced by radio continuum observations. Supplemental UV-data obtained with the OM-telescope at 210\,nm show Diffuse Ionized Gas to be well associated with UV emission originating in the underlying disk. Beside NGC\,891, NGC\,4634 is the second non-starburst galaxy with a diffuse soft X-ray halo ($|z|\le 4$\,kpc). In case of NGC\,3877, for which we also present the first high resolution X-ray imaging data, no halo emission is detectable.  {\small EPIC} pn spectra (0.3\,--\,12\,keV) of the diffuse X-ray emission are extracted at different offset positions from the disk, giving evidence to a significant decrease of gas temperatures, electron densities, and gas masses with increasing distance to the plane. A comparison between dynamical and radiative cooling time scales implies that the outflow in all targets is likely to be sustained. We find very strong indications that spatially correlated multi-phase gaseous halos are created by star forming activity in the disk plane. In a forthcoming paper, we will present multi-frequency luminosity relations and evaluate key parameters which might trigger the formation of multi-phase galaxy halos. 

}

\titlerunning{The multi-phase gaseous halos of star forming late-type galaxies}
   \maketitle
%
\section{Introduction}
For a consistent picture of the evolution and creation of multi-phase gaseous halos in star forming spiral galaxies, the structure and the physical processes in the ISM, in particular its individual components and their influence on star formation, needs to be understood. 
This is not possible without knowledge of the detailed mechanisms by which star formation processes transfer energy into the ambient medium. 
The energy input into the ISM is determined by phenomena related to young massive stars, such as supernovae (SNe), strong stellar winds or stellar photons. As a result of these processes, the ISM in the disk can connect to the halo and establish the so called disk-halo interaction \citep[e.g.,][]{De92,Br94,Da95b,Vei95,De04} with the possibility of extraplanar star formation \citep{Tu03}. The disk-halo interaction is described by theorists by means of galactic fountains \citep{Sh76,Av05}, chimneys \citep{No89}, and galactic winds \citep{Bs91, BS99}. Observationally this interaction is traced by supershells \citep{Hei84} or worm structures \citep{Ko92}. 
Possible models trying to explain gaseous galaxy halos as a consequence of the stellar feedback therefore depend on many factors, such as supernova rates, galaxy mass, magnetic fields, and the $|z|$--structure of the ISM.

Independent evidence of an interstellar disk-halo interaction comes from the presence of a thick extraplanar layer of ionized hydrogen (1.5\,--\,10\,kpc), called ``Diffuse Ionized Gas'' (DIG). This gas is assumed to be blown out into the halo by correlated SNe and was first discovered in the Milky Way \citep{Rey73} and since then in several other galaxies, such as NGC\,891 \citep{De90,Ra90}. As new Monte Carlo simulations have shown, pure photoionization by OB stars can explain most of the observed ionization structure of the DIG, assuming no or only little extra-heating \citep{Tu02,Wo04,Tu05}.
  
Very recently, significant progress in understanding the DIG was achieved, e.g., by carrying out a comprehensive H$\alpha$ survey of 74 edge-on spirals, by estimating the ejected DIG mass, and deriving an empirical set of parameters which indicates the presence of prominent DIG-halos (Rossa \& Dettmar 2000; Rossa \& Dettmar 2003a,b). 

Galaxies with nuclear starbursts (e.g., \object{NGC\,253} and \object{M\,82}), henceforth simply called starburst galaxies, are long known to possess, in addition to extended DIG layers, bubbles of hot ionized gas and X-ray plumes (e.g., \citealt{Mc95} and e.g., \citealt{Sc92,Pi00,St00,Pi01}, respectively). Contrary to starburst galaxies clear evidence of the predicted X-ray gas in halos of actively star forming spiral galaxies is, however, still scarce, although ROSAT and Chandra contributed significantly to the field with studies of the edge-on galaxy NGC\,891 \citep{Br94,Br97,St04a}. 

Supplemental studies carried out in the radio continuum have shown that the halos of starburst galaxies are populated by cosmic rays (CRs) and contain ordered magnetic fields which are oriented perpendicular to the disk \citep[e.g.,][]{Hu89,Br94,Ir99,Tu00,Da01}. These results also apply to the non-starburst case of NGC\,891 \citep{Hu91a,Hu91b}.

Finally, several studies show (or propose) correlations between H$\alpha$, radio continuum, FIR, and X-ray luminosities \citep[e.g.,][]{dej85,Co92,Wa01,Re01,Rana03}. It was argued that the presence of all these various components of the ISM in galaxy halos is related to the star formation rate (SFR) or the energy input of SNe into the ISM \citep{De92,Da95b,Ra96,Ro03a}.
In other words the spatial correlations between X-ray, radio, and H$\alpha$ (DIG) halos and those of the corresponding luminosities could simply be a consequence of stellar feedback. 

\begin{table*}
\begin{center}
\caption{Journal of observations}
\label{tab1}
\begin{tabular}{l c c c c c l l r c}
\hline
\hline
\noalign{\smallskip}
Galaxy    & $\alpha$ (J2000.0)                   & $l_{\rm II}$ & Obs. Id.   & Date      & Rev.       & Instrument & Filter & $t_{\rm int}$ & $t_{\rm GTI}$ \\
          & $\delta$ (J2000.0)                   & $b_{\rm II}$ &            & [yyyy-mm-dd]  &            &            &        &     [ks]      &   [ks]   \\
&(1)&(2)&(3)&(4)&(5)&(6)&(7)&(8)&(9)\\
\noalign{\smallskip}
\hline                                                   
\object{NGC\,891}  & 02$^{h}$22$^{m}$32\fs90       & 140.383075   & 0112280101 & $2002|08|22$ & 0495       & EPIC pn    & thin   &  15.0         &       \\
          & +42\degr20\arcmin45\farcs 8   & $-$17.417367 &            &              &            & EPIC MOS   & thin   &  18.0         &  8.2  \\
          &                               &              &            &              &            & OM         & UVW1   &  1.0          &       \\
\object{NGC\,3044}$^{a}$ & 09$^{h}$53$^{m}$41\fs03& 236.197809   & 0070940401 & $2001|11|24$ & 0359       & EPIC pn    & thin   &  18.2         &       \\
          & +01\degr34\arcmin45\farcs5    & 40.374368    & 0070940401 & $2002|05|10$ & 0443       & EPIC MOS   & thin   &  26.2         &  17.9 \\
          &                               &              &            &              &            & OM         & UVW2   &  2.9          &       \\
\object{NGC\,3221} & 10$^{h}$22$^{m}$20\fs19       & 213.973527   & 0202730101 & $2004|05|30$ & 0819       & EPIC pn    & thin   &  37.0         &       \\
          & +21\degr34\arcmin11\farcs6    & 55.713890    &            &              &            & EPIC MOS   & medium &  37.8         &  34.0 \\
          &                               &              &            &              &            & OM         & UVW2   &  5.0          &       \\
\object{NGC\,3628} & 11$^{h}$20$^{m}$16\fs93       & 240.855566   & 0110980101 & $2000|11|27$ & 0178       & EPIC pn    & thin   &  50.8         &       \\
          & +13\degr35\arcmin13\farcs7    & 64.779168    &            &              &            & EPIC MOS   & thin   &  54.5         &  48.7 \\
          &                               &              &            &              &            & OM         & UVW2   &  2.0          &       \\
\object{NGC\,3877} & 11$^{h}$46$^{m}$07\fs81       & 150.719159   & 0047540701 & $2001|06|17$ & 0279       & EPIC pn    & thin   &  23.8         &       \\
          & +47\degr29\arcmin40\farcs9    & 65.956495    &            &              &            & EPIC MOS   & thin   &  27.7         &  14.1 \\
          &                               &              &            &              &            & OM         & n. a.  &               &       \\
\object{NGC\,4631} & 12$^{h}$42$^{m}$07\fs79       & 142.814930   & 0110900201 & $2002|06|28$ & 0467       & EPIC pn    & thin   &  50.8         &       \\
          & +32\degr32\arcmin26\farcs9    &  84.223809   &            &              &            & EPIC MOS   & medium &  53.8         &  44.8 \\
          &                               &              &            &              &            & OM         & UVW2   &  1.0          &       \\
\object{NGC\,4634} & 12$^{h}$42$^{m}$40\fs80       & 293.454159   & 0202730301 & $2004|06|13$ & 0826       & EPIC pn    & thin   &  37.0         &       \\
          & +14\degr17\arcmin49\farcs1    & 77.007282    &            &              &            & EPIC MOS   & medium &  38.7         &  34.7 \\
          &                               &              &            &              &            & OM         & UVW2   &   4.5         &       \\
\object{NGC\,4666} & 12$^{h}$45$^{m}$08\fs35       & 299.535358   & 0110980201 & $2002|06|27$ & 0467       & EPIC pn    & thin   &  54.6         &       \\
          & $-$00\degr27\arcmin48\farcs6  & 62.366702    &            &              &            & EPIC MOS   & thin   &  58.0         &  55.6 \\
          &                               &              &            &              &            & OM         & UVW2   &  4.0          &       \\
\object{NGC\,5775} & 14$^{h}$53$^{m}$57\fs48       & 359.430954   & 0150350101 & $2003|07|28$ & 0665       & EPIC pn    & thin   &  41.1         &       \\
          & +03\degr32\arcmin39\farcs5    & 52.423232    &            &              &            & EPIC MOS   & thin   &  41.4         &  23.4 \\
          &                               &              &            &              &            & OM         & UVW2   &  5.0          &       \\
\hline
\noalign{\smallskip}
\end{tabular}
\end{center}
\vspace{-0.3cm}
{\small Notes:\quad Cols. (1) and (2): Equatorial and Galactic coordinates of the targets. Col. (3): XMM-Newton observation identification number. Col. (4): Date of observation. Col. (5): Revolution number. Cols. (6) and (7): Instrument and filter used during observations. In some cases 'medium' blocking filters had to be used to avoid saturation of the EPIC MOS detectors. Col. (8): Total integration time. Col. (9): GTI-corrected integration times, determined from merged EPIC (pn and MOS) observations.

{\small ($a$)} Integration times for NGC\,3044 are listed for revolution number 0443. $t_{\rm GTI}$ has been determined from both runs.}
\end{table*}

\begin{table*}
\begin{center}
\caption{Basic parameters of the sample}
\label{tab2}
\begin{tabular}{l c c c c c c c c c c}
\hline
\hline
\noalign{\smallskip}
Galaxy    & Type   & D     & $i$     & $N_{\rm H}$            & $v_{\rm \ion{H}{i}}$            & $S_{60}/S_{100}$ & $L_{\rm FIR}/D^2_{25}$               & \multicolumn{3}{c}{halo-type}     \\ 
          &        & [Mpc] & [\degr] & [10$^{20}$\,cm$^{-2}$] & [km s$^{-1}$] &                  & $[10^{40}$\,erg s$^{-1}$ kpc$^{-2}$] & DIG       & Radio     & X-ray     \\
\noalign{\smallskip}
\hline
NGC\,891  & Sb     &  9.5  & 88                 & 7.64      &  528                   & 0.3869           &  6.92                                & $\bullet$ & $\bullet$ & $\bullet$ \\
NGC\,3044 & Sc     & 17.2  & 84                 & 3.53      & 1292                   & 0.4974           &  8.52                                & $\bullet$ & $\bullet$ & $\bullet$ \\
NGC\,3221 & SBcd   & 54.8  & 77                 & 2.03      & 4110                   & 0.4115           &  13.5                                & $\bullet$ & $\bullet$ & $\bullet$ \\
NGC\,3628 & Sb     & 10.0  & 80                 & 2.23      &  847                   & 0.5182           &  8.24                                & $\bullet$ & $\bullet$ & $\bullet$ \\
NGC\,3877 & Sc     & 12.1  & 83                 & 2.22      &  904                   & 0.3443           &  5.14                                & $\circ$ & $\circ$ & $\circ$   \\
NGC\,4631 & Sc     &  7.5  & 86                 & 1.27      &  606                   & 0.5334           &  7.76                                & $\bullet$ & $\bullet$ & $\bullet$ \\
NGC\,4634 & Sc     & 19.1  & 83                 & 2.24      &  118                   & 0.3720           &  12.4                                & $\bullet$ & $\bullet $          & $\bullet$ \\
NGC\,4666 & Sc     & 20.2  & 80                 & 1.74      & 1520                   & 0.4318           &  31.9                                & $\bullet$ & $\bullet$ & $\bullet$ \\
NGC\,5775 & Sc     & 26.7  & 84                 & 3.48      & 1681                   & 0.4240           &  24.6                                & $\bullet$ & $\bullet$ & $\bullet$ \\
\hline
\end{tabular}
\end{center}
\vspace{-0.2cm}
{\small Notes:\quad Total Galactic \ion{H}{i} column densities were taken from \citet{DL90}. These values are considered to be representative column densities, as the \ion{H}{i} beam sizes (1\arcmin -- 20\arcmin) are comparable with the optical $D_{25}$ diameters of our targets (2\farcm6 -- 14\arcmin). Revised IRAS flux ratios have been adopted from \citet{Sa03}. For $L_{\rm FIR}/D^{2}_{25}$, we used the revised $S_{60}/S_{100}$ ratios and the optical diameters of the 25$^{\rm th}$ magnitude isophote ($D_{25}$) given by \citet{rc3}. All other values are from \citet{Ro00,Ro03a}. \ion{H}{i} velocities are mean heliocentric neutral hydrogen velocities taken from RC3 \citep{rc3}. The meaning of the symbols is as follows: $\bullet$ = detections, $\circ$ = non-detections.}
\end{table*}

For starburst galaxies,
correlations between the presence of DIG and soft X-ray halos as well as a dependence between diffuse X-ray luminosities and the energy input rate from massive stars into the ISM could already be established, making extensive use of the Chandra satellite (Strickland et al. 2004a,b). For actively star forming (non-starburst) spiral galaxies all these correlations still need to be investigated.

Recent cosmological models, however, claim a completely different mechanism to be responsible for the creation of extended galaxy halos, namely accretion from the IGM with subsequent infall of hot gas from the halo onto the disk plane \citep{Be00,To02}.

In addition to the morphological correlations between H$\alpha$, radio continuum, and diffuse X-ray halos, the present study intends to trace correlations between radio continuum, FIR, and X-ray luminosities down to lower energy input rates (i.e. SFRs). We will consider additional wavelength regions (UV, B, and H$\alpha$) and include results from starburst \citep{St04a,St04b} and actively star forming galaxies, to investigate whether the multi-frequency correlations also hold at other wavelengths and for a larger variety of SFRs (T\"ullmann et al. in prep.).

Some of the basic questions we want to address in this study are the following: How common are multi-phase gaseous halos among actively star forming (non-starburst) galaxies?
Is there a relation between the individual gaseous halo components of actively star forming spiral galaxies (e.g., similar to those considered for starburst galaxies)? What parameters determine the presence of large scale halos in starburst and normal star forming galaxies? Can we constrain their formation, e.g., by distinguishing between the outflow and the infall scenario and is there an energy input threshold above which multi-phase halos start to form? 

In this paper, the inital part of the study, we first concentrate on the presentation of new XMM-Newton observations of the diffuse Hot Ionized Medium (HIM) in nine star forming edge-on spiral galaxies. Despite better spatial resolution of the Chandra X-ray observatory, the substantially larger effective mirror area and its high sensitivity to detect faint extended emission make XMM-Newton the preferable instrument. 

After introducing the sample and the main selection criteria (Sect. 2), and outlining the basic steps of the data reduction (Sect. 3), we discuss the main results presented in this work (Sect. 4). Our data consist of merged {\small EPIC} (European Photon Imaging Camera) pn and MOS contour maps (0.2\,--12.0\,keV) overlaid onto H$\alpha$ images (Subsect. 4.1) and Optical Monitor (OM) imagery with the UVW2(1) filter (Subsect. 4.2). In Subsect. 4.3 we show {\small EPIC} pn spectra (0.3\,--12.0\,keV) of the disks and halos of our targets followed by an investigation of the up to now spatially best resolved temperature gradient of the HIM. Important plasma parameters are calculated, such as electron densities, radiative cooling times, and gas masses. Finally, Sects. 5 and 6 briefly summarize our understanding of the formation of multi-phase gaseous halos and outline the main results and conclusions of this work.  

The follow-up paper will relate our results presented here to multi-wavelength data obtained in the FIR, radio, H$\alpha$, and B-band, in order to investigate the expected correlation among the individual halo components for normal star forming and starburst galaxies. Special preference will be given to conceive a consistent picture on the formation of multi-phase gaseous galaxy halos.  

\section{The sample}
The sample was chosen from the aforementioned H$\alpha$ survey of edge-on galaxies \citep{Ro03a} by selecting galaxies with the highest $L_{\rm FIR}/D_{25}^{2}$ values and exceeding a IRAS $S_{60}/S_{100}$ ratio of 0.3. Usually these parameters are considered to be a good discriminator for starburst and normal star forming spiral galaxies \citep[e.g.,][]{lehe,Ro03a,St04a}. Moreover, these ratios seem to be sensitive to the existence of extended DIG and radio continuum halos. 

Consequently, our initial sample is still biased against lower star formation rates, which means that starburst galaxies dominante the sample. In order to address the above mentioned issues, e.g., to investigate a possible energy input threshold above which multi-phase halos are created, galaxies with lower $S_{60}/S_{100}$ and $L_{\rm FIR}/D_{25}^{2}$ ratios need to be considered. This will be subject of the follow-up paper.

All observations of the current sample were carried out with XMM-Newton. Four galaxies, namely NGC\,3044, NGC\,3221, NGC\,4634, and NGC\,5775, have been observed during AO-02/03. 
The remaining ones have been collected from the XMM-Newton data archive. Two of them, NGC\,891 and NGC\,3877, show by far the lowest $S_{60}/S_{100}$ and $L_{\rm FIR}/D_{25}^{2}$ ratios among our sample. Tables~\ref{tab1} and \ref{tab2} present a journal of observations and basic properties of the sample, respectively. For individual notes on the targets, see \citet{Ro00,Ro03b}.  

We like to emphasize that the current work is the initial part of our study and that the presented sample is neither complete nor free from selection effects.
 
\section{XMM-Newton observations and data reduction}
All targets have been observed with {\small EPIC} pn \citep{Str01} and {\small EPIC} MOS \citep{Tur01} as well as with the OM telescope \citep{Ma01} on-board the XMM-Newton satellite\footnote{Instrument specific information are provided by the XMM-Newton User's Handbook and are accessible through this link: http://xmm.vilspa.esa.es/external/xmm\_user\_support/docu-\\mentation/uhb/index.html}. Data for NGC\,3044 were collected twice, as during the first run in November 2001 a solar flare prevented the entire execution of the observation. 
Unfortunately, the collected data during the second run in May 2002 could also not be used completely as it was seriously affected by a high radiation background. In case of NGC\,5775 there are periods of a strongly enhanced background as well, but less pronounced as for NGC\,3044. The same applies to NGC\,891. 

Good time intervals (GTIs) of the {\small EPIC} eventlists, exposure times $t_\mathrm{int}$, filters and instruments, revolution numbers, observing dates, observation identification numbers, and the coordinates of the sample galaxies are listed in the journal of observations (Table~\ref{tab1}). 
Data reduction for pn, MOS, and OM frames was carried out with standard tasks provided by SAS v6.0.0 using the pipeline scripts {\small EPCHAIN}, {\small EMCHAIN}, and {\small OMICHAIN}. 
All {\small EPIC} observations were carried out in full frame mode using the thin or the medium blocking filter. The OM telescope was operated in default imaging mode using the UVW1 and UVW2 filters. 

After first inspection of the pn and MOS data, light curves were produced to check for possible flaring events. As in almost all frames significant high background periods were found, individual GTI corrections were applied (cf. Table~\ref{tab1}). 

Six different energy bands are used for the analysis: 0.2\,--0.5\,keV (supersoft), 0.5\,--\,1.0\,keV (soft), 1.0\,--\,2.0\,keV (medium), 2.0\,--\,4.5\,keV (medium/hard), 4.5\,--\,12.0\,keV (hard), and 0.2\,--\,12\,keV (total, as sum of the previous bands).
All {\small EPIC} pn eventlists were filtered to remove hot and bad pixels as well as to lower electronic noise by selecting {\small FLAG}=0 and {\small PATTERN}=0 for supersoft energies and {\small FLAG}=0 and {\small PATTERN} $\le4$ for the other bands. In case of MOS we used {\small FLAG}=0 and {\small PATTERN} $\le12$ (accepting ``singles'' to ``quadruples'').

From screened eventlists cleaned images were produced in the different energy bands, excluding regions of significant background variability \citep[see][]{Pi04}.  
In order to be sensitive down to energies of 0.2\,keV, noisy pixels of the pn and MOS CCDs were rejected. 

A correction for out-of-time events has been applied to all pn frames. In a final step, all pn and MOS images of each energy band were merged and smoothed using a Gaussian filter technique with a FWHM of 20\arcsec.

As we are mainly interested in the diffuse X-ray emission, we cleaned the {\small EPIC} pn eventlist for the spectral analysis from unresolved point sources and other contaminating objects around our targets using circular apertures. A point source subtracted image has been created to check whether all sources have been masked out successfully.
Due to the higher quantum efficiency and larger effective area at low energies only {\small EPIC} pn rather than MOS-eventlists have been used for spectral analysis. Spectra have been extracted from the disk plane and from different offset positions in the halo, considering only single and double events ({\small PATTERN} $\le4$). 

For each science and background region, the extracted area has been determined with the {\small BACKSCALE} task. Subsequently, response matrices (task {\small RMFGEN}) together with corresponding detector maps and auxiliary response files (task {\small ARFGEN}) have been created. As the linear spatial extent of all targets is $\le 17\arcmin$, there are at least 4 areas which are usable for background subtraction. In order to achieve reasonable S/N statistics, these local backgrounds have to be located at comparable offsets to the readout node as the extended emission and if possible on the same or neighboring CCD. Each background was then subtracted from a halo spectrum to find the background frame best suited for subtraction.

Finally, all science spectra were background subtracted and grouped with typically 40\,--\,180 counts per bin for the disk and about 55\,--\,330 for the halo. The convolution with the instrumental response and the spectral fitting was carried out with XSPEC (v11.2.0). 
  
The UV-data, obtained with the Optical Monitor, were pipeline processed and calibrated using the SAS-task {\small OMICHAIN}. For the analysis we used the UVW1 and UVW2 filters, because only in these wave bands the count rates of bright point sources outside the field of view are low enough so that parasitic light, as for instance visualized by straylight ellipses, ring structures or ghosts, are of no concern. The UVW2 filter covers the approximate wavelength range from 180\,nm to 225\,nm\ and peaks at about 210\,nm. Unfortunately, OM observations of NGC\,891 were not carried out with the UVW2 filter (archive data). Instead the UVW1 filter was used which has its peak transmission at about 270\,nm and covers the range between 245\,nm and 320\,nm. The PSFs (FWHM) of the UVW1 and UVW2-filter amount to $1\farcs7$ and $2\farcs3$, respectively. Integration times for the selected UV-filters are listed in Table~\ref{tab1}. 
\section{Results and discussion}
There is obviously a tight linear relation between FIR radiation and the radio continuum emitted by star forming spiral galaxies \citep[e.g.,][]{dej85,Hel85,Re01,Co02}. This correlation can be explained if one considers the origin of the radiation. FIR emission is mainly produced in star forming regions by young dust enshrouded stars whose UV flux is absorbed by dust grains and re-emitted at infrared wavelengths. Extended synchrotron halos are seen in the radio continuum and trace high energy electrons produced by SNe or supernova remnants (SNRs). Apparently, both kinds of radiation trace high mass star forming processes in the disk and hence can be related to typical parameters, such as the SFR \citep{Co92,ken98}.

It appears likely that this relation also holds if additional star formation tracers are included, such as B-band, H$\alpha$, UV, and X-ray luminosities.
H$\alpha$ is known to be a good tracer of gas photoionized by OB-stars, such as the halo DIG \citep[e.g.,][]{Rey84,Tude00,Wo04,Tu05}. X-ray emission in galaxies typically originates from X-ray binaries or SNRs but also from superwinds and extraplanar diffuse hot plasma generated by these objects. The others, finally, are reliable indicators of the continuum radiation emitted by young and hot stars. 
Thus, we expect a multi-frequency correlation between DIG, HIM, and CRs, which should hold for a large bandwidth of SFRs, from the quiescent to the most powerful starburst cases.

For most of the targets H$\alpha$, IRAS, and VLA radio continuum data exist and have already been published. Hence, the final step aims at the investigation of the still poorly constrained relationship at X-ray wavelengths. 
In the following, XMM-Newton {\small EPIC} and OM observations of the HIM in halos of nine star forming galaxies are presented and compared to the DIG-halos visible in H$\alpha$. 

It is a remarkable observational result that all galaxies, except NGC\,3877, possess extended soft X-ray, DIG, and radio continuum halos and that extraplanar DIG is obviously well associated with UV emitting sources located in the underlying disk.

\subsection{EPIC pn and MOS imagery}
Throughout Figs.~\ref{F1}\,--\,\ref{F10} we provide contour plots of the observed X-ray emission covering the supersoft (0.2\,--\,0.5\,keV, upper left panel), soft (0.5\,--\,1.0\,keV, upper right panel), medium (1.0\,--\,2.0\,keV, middle left panel), medium/hard (2.0\,--\,4.5\,keV, middle right panel), hard (4.5\,--\,12.0\,keV, lower left panel), and the combined (0.2\,--\,12.0\,keV, lower right panel) energy bands. The contour plots are overlaid onto corresponding H$\alpha$ images from \citet{Ro00,Ro03a}, unless otherwise noted. Contours always start at $3\,\sigma$ above the local background and are displayed on a square-root intensity scale with a spacing of $\sqrt{2}$.

\begin{figure*}
\hspace{.25cm}
\caption{Merged {\small EPIC} (pn and MOS) X-ray contours of NGC\,891 (square-root intensity scale) between $7.4\times 10^{-6}\,{\rm cts\ s^{-1}\,pix^{-1}}$ and $2.0\times 10^{-5}\,{\rm cts\ s^{-1}\,pix^{-1}}$ (at a contour smoothness of 4 in ds9) and overlayed onto an H$\alpha$ image \citep{Ro03b}. The emission at supersoft energies is affected by foreground absorption. A linear length scale of 1\arcmin\ corresponds to 2.8\,kpc. The position of SN1986J is indicated by a white cross; the HST blow-up \citep{Ro04} shows only a small part of NGC\,891.}
\label{F1} 
\end{figure*}

\subsubsection{NGC\,891}
NGC\,891 has often been named 'twin' of the Milky Way regarding its Hubble-type, galaxy mass, and bolometric luminosity \citep{vk84} and is so far the only non-starburst galaxy with a positive detection of a gaseous X-ray halo \citep{Br94,Br97,St04a}.

From our XMM-Newton observations, shown in Fig.~\ref{F1}, we also confirm the existence of an extended diffuse gaseous X-ray halo, which is most pronounced in the soft energy band (0.5\,--\,1.0\,keV). At supersoft energies (0.2\,--\,0.5\,keV) merely point sources and little extended diffuse emission is visible, indicating that the supersoft X-ray emission is absorbed by heavy metals along the line of sight. Actually, the total galactic H{\small I} column density of $N_{\rm H} =7.64\times 10^{20}$\,cm$^{-2}$ reported by \citet{DL90} turns out to be the largest among our sample, and hence, our observations seem indeed to be strongly affected by foreground absorption. Moreover, there are also clues to strong internal absorption effects, as confirmed by the lack of emission along the major axis close to the center of the galaxy (cf. upper left panel) and by a prominent dust layer reported by \citet{HS97}. 

At medium energies between 2.0 and 4.5\,keV, there is still significant emission, likely from point sources, such as SNRs or X-ray binaries, but also from extended diffuse extraplanar gas. In the 4.5\,--\,12.0\,keV band, only hard point sources in the disk are visible and the inner 4\,kpc resolve at least into two discrete sources. The southern source at $\alpha$\,(J2000)\,=\,$02^{\rm h}22^{\rm m}31\fs33$ and $\delta$\,(J2000)\,=\,$42\degr19\arcmin56\farcs4$ can be identified as SN1986J \citep{Br94,Ba99}; the one located at $\alpha$\,(J2000)\,=\,$02^{\rm h}22^{\rm m}31\fs30$ and $\delta$\,(J2000)\,=\,$42\degr20\arcmin30\farcs0$ turns out to be the brightest X-ray source of NGC\,891 for energies $>$\,0.5\,keV. The nature of this source remains to be investigated by means of further spectroscopy. Both sources near the center are also visible on the hard (2.0\,--\,8.0\,keV) Chandra X-ray image presented by \citet{St04a}. 

Star formation, as traced by \ion{H}{ii}-regions in the disk, appears to be enhanced in the NE part and in the center of NGC\,891. This can be seen in more detail on the HST H$\alpha$ blow-up from \citet{Ro04} which is shown in the upper left panel of Fig.~\ref{F1}. Wherever the H$\alpha$ emission is strong, extraplanar DIG and soft diffuse X-rays are well aligned. As soft X-rays and emission from DIG are both affected by internal absorption, this effect is considered to be less important for the correlation, especially at extraplanar distances. The spatial coincidence between DIG and soft X-rays strongly suggests that both gaseous components are produced by star forming activity in the disk plane rather than by accretion from the IGM.
An inhomogeneous and filamentary distribution of the ionized gas was found for NGC\,891 by H$\alpha$ imaging from the ground \citep{De90,Ra90} and with HST/WFPC2 \citep{Ro04}, as well as the presence of a prominent radio continuum halo which also reaches the largest $|z|$-extent above the NE part of the disk \citep[e.g.,][]{Da94}. The global and local spatial correlations of the different ISM constituents in the halo with star formation (SF) in the disk support the origin of gaseous halos by the disk-halo interaction.

Moreover, the existence of \ion{H}{ii}-regions located several hundred pc above the disk plane \citep{HS97,Ro04} directly implies the possibility of extraplanar star formation. As the lifetime of the central stars is usually too low compared to the traveltime needed to reach the halo, these objects must have been formed from gas deposited in the halo by an earlier SF event \citep{Tu03}. Therefore, extraplanar star formation provides independent support for the disk-halo interaction.

The diffuse X-ray emission is most extended above the nucleus and reaches its largest detected extent of $z$\,=\,5.6\,kpc in the NW part of the halo (3.2\,kpc in the SE part). If one accounts for emission originating from the extragalactic point source located at $\alpha$\,(J2000)\,=\,$02^{\rm h}22^{\rm m}24\fs37$ and $\delta$\,(J2000)\,=\,$42\degr21\arcmin39\farcs0$ \citep{Har03}, the extent of the northern halo still would be $\approx 3$\,kpc. However, from the point-source-extracted soft X-ray image presented by \citet{St04a}, diffuse extraplanar emission can be established on scales of $\approx$\,6.5\,kpc. This, as well as the detection of a type-II SN, also hints at violent star formation in the inner part of that galaxy. 
As the GTI-correction applied to the data cut down the total integration time to 8.2\,ks, we expect the size of the halo of NGC\,891 to be significantly larger.
\begin{figure*}
\caption{{\small EPIC} overlays for NGC\,3044. Contours are spaced by factors of $\sqrt2$ between $5.3\times 10^{-6}\,{\rm cts\ s^{-1}\,pix^{-1}}$ and $1.5\times 10^{-5}\,{\rm cts\ s^{-1}\,pix^{-1}}$ (contour smoothness = 4) and overlaid onto the H$\alpha$ image presented by \citet{Ro00}. A substantial fraction of photons at supersoft energies is absorbed (cf. Table~\ref{tab2}). The position of SN1983E is marked in the middle left panel. The H$\alpha$ blow-up in the upper left panel shows that DIG and \ion{H}{ii}-regions are well associated. 1\arcmin\ corresponds to 5.0\,kpc.}
\label{F2} 
\end{figure*}

\begin{figure*}
\hspace{0.5cm}
\caption{Merged {\small EPIC} X-ray contours overlaid onto the H$\alpha$ image of NGC\,3221 \citep{Ro03b}. Contours (square-root intensity scale) are spaced by factors of $\sqrt2$ between $3.8\times 10^{-6}\,{\rm cts\ s^{-1}\,pix^{-1}}$ and $4.0\times 10^{-4}\,{\rm cts\ s^{-1}\,pix^{-1}}$ (contour smoothness = 3). In case of NGC\,3221, 1\arcmin\ amounts to 16\,kpc. Note the position of SN1961L at the rim of the northern spiral arm. The bar and the DIG features associated with the ends of the bar are marked on the magnified H$\alpha$ image in the upper left subpanel.}
\label{F3} 
\end{figure*}

\subsubsection{NGC\,3044}
This target, sometimes classified as ``starburst'' galaxy, has been studied by optical photometry \citep{Ro00,Co00,Mi03a} and spectroscopy \citep{Tude00} with clear detections of a DIG-halo reaching up to 4\,kpc off the disk plane. NGC\,3044 has a radio continuum halo of similar size \citep{Hu89,Ir00} and is also listed in the IRAS bright galaxy sample \citep{Soifer}, indicating enhanced star formation in the disk. 

Currently, it is not completely clear whether or not NGC\,3044 is a starburst galaxy. Although the high $S_{60}/S_{100}$ ratio, the enhanced IR luminosity, and the outflow features visible on the H$\alpha$ image presented by \citet{Co00} would indeed imply NGC\,3044 being a starburst galaxy, there are two important aspects which might cast doubts on that classification. Firstly, beside a high $S_{60}/S_{100}$ ratio, a second measure which we adopt here to distinguish between the (nucleated) starburst and non-starburst case, is the relation between IR luminosity and the effective H$\alpha$ half-light radius \citep{lehe}. According to their Fig.~2, NGC\,3044 lies towards the bottom right of that plot, a region which is populated by normal (non-starburst) galaxies. Secondly, diagnostic diagrams of the central part of NGC\,3044 characterize the gas to be mainly photoionized but with weak contributions from shocks and with only mediocre line broadening \citep{Tude00}. 

On the other hand, both arguments would also be consistent with a more extended starburst. In fact, NGC\,3044 is as far away from the dashed line in Fig.~2 of \citet{lehe} as the starburst galaxy NGC\,5775, which has a comparable $S_{60}/S_{100}$ ratio and a slightly higher IR-luminosity (see below). Taking all this together, we consider NGC\,3044 to be a starburst galaxy with extended star formation over a large fraction of the disk (see \citet{Ro00} and \citet{Co00} for detailed H$\alpha$ images). As a consequence, the starburst in NGC\,3044 appears to be different from the so called ``nucleated'' starbursts, seen in NGC\,3628, M82, or NGC\,253.

In Fig.~\ref{F2} we present the first X-ray observations of NGC\,3044 which show an extended soft X-ray halo on typical scales of about 3\,kpc, but no elongated outflow features which are often seen in nucleated starburst galaxies such as M82 or NGC\,253. 
Similar to NGC\,891, star formation is not uniformly distributed as proved by the depression in the H$\alpha$ distribution of the eastern disk. A similar depression is seen in the \ion{H}{i} column density distribution of NGC\,3044 \citep{Le97}, suggesting that internal absorption is of minor importance. Again, the H$\alpha$ image shown in the upper left subpanel of Fig.¸~\ref{F2} \citep[][logarithmic intensity scale]{Ro00} evidences that extraplanar DIG and \ion{H}{ii}-regions are closely related. As $41\%$ of the overall H$\alpha$ flux originates in the halo of NGC\,3044 \citep{Ro00}, there are strong indications that a substantial fraction of the DIG is extraplanar.

The soft energy band (0.5\,--\,1.0\,keV) confirms again a good spatial correlation between DIG and X-ray emission. Both gas phases reach their maximum $|z|$-extent where star formation is most pronounced, namely in the innermost 5\,kpc of the disk. The same also holds for the radio continuum halo observed at 1.5\,GHz by \citet{Hu89}. 

The emission in the supersoft band is again strongly affected by foreground absorption from the Milky Way and indicates only few point-like sources, two of which are best visible at high energies $>$\,4.5\,keV. Interestingly, the eastern source in the hard band can be associated with SN1983E \citep[][see also middle left panel of Fig.~\ref{F2}]{Ba99}, indicating a site of recent star forming activity. As this site seems to coincide with \ion{H}{ii}-regions \citep[see also][]{Ro00} which are associated with extraplanar DIG, the existence of a CR-halo which is produced by multiple SNe and SNRs consistently fits into the picture of star formation induced galaxy halos. 

It should be stressed that the point source above the center of the northern disk plane, seen in the medium/hard and combined energy bands, could be due to a background galaxy cluster whereas the easternmost X-ray source visible at soft and intermediate energies can be associated with a cluster of \ion{H}{ii}-regions located at the far end of the spiral arm.  
\subsubsection{NGC\,3221}
According to its high $S_{60}/S_{100}$ ratio of 0.42 and a strongly increased $L_{\rm FIR}/D^{2}_{25}$ ratio of $13.5\times10^{40}$\,erg s$^{-1}$\,kpc$^{-2}$, NGC\,3221 could also be considered to be a starburst galaxy. These ratios clearly suggest an enhanced SFR as well as the presence of extraplanar DIG, radio continuum, and X-ray emission. Previous H$\alpha$ \citep{Ro03b} and radio continuum observations \citep{Ir99} give indeed evidence of vertically extended disk emission in these wave bands. However, the rather low inclination of $i=77^\circ$ hampers a clear discrimination of radiation originating in or off the disk. Therefore, it is more reliable to talk about ``extraplanar'' emission than about ``halos'' which imply extended structures of several kpc in size. 

In Fig.~\ref{F3} we show the first X-ray imaging data obtained for NGC\,3221. X-ray contours of the soft, medium, and hard energy bands are overlaid onto an H$\alpha$ image and nicely show the soft diffuse X-ray emission (0.5\,--\,1.0\,keV) leaking off the disk. All overlays have been smoothed to a resolution of 20\arcsec, i.e. structures, such as the point source visible in the upper right corner of the soft energy band are clearly resolved. If we correct for projection effects, we determine the diffuse soft X-ray emission to extend up to 8\,kpc above the center of NGC\,3221. 

Except in the medium energy band, X-rays are lacking completely in the southern part, whereas the northern hemisphere is X-ray-bright throughout all energies. 
There is no supersoft emission coincident with the northern spiral arm which could be due to foreground and internal absorption or due to the lack of X-rays. Both ends of the bar show enhanced soft X-ray and DIG emission (see marks in the magnified H$\alpha$ image shown in the upper left subpanel of Fig.~\ref{F3}).
    
As can be seen in more detail on the H$\alpha$ image presented by \citep{Ro03b}, the northern hemisphere also shows strong star formation and the presence of extraplanar DIG. The detection of a SN several decades ago \citep[SN1961L,][]{Ba99} located approximately 10$\arcsec$ above the northern edge of the H$\alpha$-disk (see label in Fig.~\ref{F3}) also provides a possible independent link to star formation activity. In case of NGC\,3221 extraplanar X-ray and DIG emission are spatially correlated and also coincide with the detected extraplanar radio continuum emission \citep{Ir99}.    
\begin{figure*}
\hspace{-1.55cm}
\caption{{\small EPIC} X-ray overlays produced in different energy bands for NGC\,3628. Contours start at $3\sigma$ or $2.8\times 10^{-6}\,{\rm cts\ s^{-1}\,pix^{-1}}$, respectively and end at $8.0\times 10^{-4}\,{\rm cts\ s^{-1}\,pix^{-1}}$. Given the large dimensions of this galaxy, a DSS-2-image (B-band) has been used rather than the H$\alpha$ image \citep[see][]{Ro03b}. The black solid lines indicate the positions of the extraplanar filamentary structures visible in H$\alpha$. In case of NGC\,3628 a linear length scale of 1\arcmin\ represents 2.9\,kpc.}
\label{F4} 
\end{figure*}

\subsubsection{NGC\,3628}
This well studied interacting object is a member of the Leo-triplet and represents a good example of a galaxy undergoing a nuclear starburst. It has been previously observed with ROSAT \citep{Da96,Re97} and Chandra \citep{St04a} as well as in the radio continuum \citep{Co87,Dum95} and FIR \citep{Ri93}. 

The X-ray data presented in Fig.~\ref{F4} show that the disk plane of NGC\,3628 is heavily obscured by a prominent dust lane, blocking most of the optical and X-ray radiation coming from the disk. Strong internal absorption is also visible in the supersoft X-ray image being responsible for the apparent dichotomy of the disk. This effect vanishes at higher energies and above 1.0\,keV a central point source appears. Actually, the X-ray emitter close to the nucleus is the strongest source throughout all energy bands and is most likely a variable high mass X-ray binary \citep{Da95a,Wa04}. 

Moreover, our data show a very elongated diffuse soft X-ray halo, apparently associated with the starburst in the nucleus. As can further be seen from these images, the filamentary DIG structures \citep[e.g.,][see also black solid lines in Fig.~\ref{F4}]{Fa90,Ro03b} seem to envelop the X-ray hot gas in the halo. It appears plausible to consider these structures to represent the outer limb brightened edges of a collimated outflow cone. Therefore, this morphology together with the detected CR-halo would give direct support for the disk-halo interaction. In addition, the configuration of the different gaseous components is typical for the superbubble breakout scenario \citep{Ma88,Ma99} and thus strengthens the hypothesis that gaseous galaxy halos are created by star forming activity in the disk.  

The X-ray overlays of the (2.0\,--\,4.5\,keV) energy band in Fig.~\ref{F4} indicates the existence of six extraplanar point sources (four in the northern and two in the southern halo) which seem to be located in the region of the outflowing gas. Most of these sources are likely to be background quasars \citep{He89,Arp02} and are also visible at soft energies ($<$\,1.0\,keV) which makes a precise determination of the overall extent of the diffuse X-ray emission difficult. 
Although we did not apply a point source extraction to these regions we can reliably trace soft, diffuse X-ray emission out to 9.5\,kpc in the northern halo. 
In principle we can put a conservative lower limit of about 8.7\,kpc on the $|z|$-extent of the southern halo. However, as the two point sources in the southern halo are well resolved, the soft diffuse emission visible between these objects could trace outflowing gas associated with the central starburst in NGC\,3628. If this assumption is correct, we detect soft X-rays at extraplanar distances of 13.3\,kpc, making this galaxy to possess one of the most extended X-ray halos known so far. 

Recently, \citet{St04a,St04b} made use of the better point source rejection capabilities of the Chandra observatory and presented images of the X-ray hot gas in NGC\,3628 which seem to be less sensitive (due to sensitivity losses of the S2 chip) than our data.  

\begin{figure*}
\hspace{1.cm}
\caption{{\small EPIC} X-ray contours for NGC\,3877 (ranging from $2.8\times 10^{-6}\,{\rm cts\ s^{-1}\,pix^{-1}}$ to $6.5\times 10^{-4}\,{\rm cts\ s^{-1}\,pix^{-1}}$). Strong geometrical distortions, introduced by the H$\alpha$-filter, did not yield a reasonable astrometric accuracy for producing overlays. A greyscale DSS2 B-band image was used instead, the H$\alpha$ blow-up is shown in the upper left panel. Note that the supersoft emission is affected by significant foreground absorption. The position of SN1998S is indicated in the lower left panel. A linear length scale of 1\arcmin\ corresponds to 3.5\,kpc.}
\label{F5} 
\end{figure*}

\begin{figure*}
\caption{{\small EPIC} X-ray overlays of NGC\,4631 created from merged pn and MOS images in different energy bands. Contours are separated by factors of $\sqrt2$ and plotted on a square-root intensity scale between $2.3\times 10^{-6}\,{\rm cts\ s^{-1}\,pix^{-1}}$ and $7.2\times 10^{-5}\,{\rm cts\ s^{-1}\,pix^{-1}}$. The total $|z|$-extent of the soft and supersoft extraplanar X-ray emission of NGC\,4631 is not entirely covered by the H$\alpha$ filter image presented by \citet{Go96}. It reaches approximately 45\arcsec\ ($\sim 1.65$\,kpc) further out into the northern halo. 1\arcmin $\approx$ 2.2\,kpc.} 
\label{F6} 
\end{figure*}
\subsubsection{NGC\,3877}
NGC\,3877 is actually a poorly studied non-starburst galaxy regarding the disk-halo interaction. As can be seen from the H$\alpha$ blow-up, shown in the upper left panel of Fig.~\ref{F5}, it appears to be very unlikely that, despite of possible projection effects, this object possesses significant extraplanar DIG, not to mention an extended H$\alpha$ halo. The 20\,cm radio continuum map presented by \citet{Co87} shows a putative halo with an average $|z|$-extent of about 4\,kpc. Given the low angular resolution of the VLA observations ($\theta\approx1\arcmin$) and taking into account that point sources have about the same size as the minor axis extent of NGC\,3877, this galaxy is most likely lacking a radio continuum halo. Moreover, the determination of the vertical extent of the radio continuum radiation is also hampered by a relatively low inclination angle of $i=83\degr$. A significant fraction of the ``extraplanar'' radio continuum is likely to be associated with the outer edges of the disk plane rather than with the halo. 
Among our sample this galaxy has with 0.34 the lowest S$_{60}$/S$_{100}$ ratio and the lowest $L_{\rm FIR}/D^{2}_{25}$ value ($5.14\times10^{40}$\,erg s$^{-1}$\,kpc$^{-2}$, cf. Tab.~2) which would also be consistent with the absence of multi-phase halos.

In this context, the X-ray data of NGC\,3877, presented in Fig.~\ref{F5}, provide a not surprising result, namely the lack of extraplanar X-ray emission.
Diffuse X-rays are only detectable to a significant amount in the soft band around the central region of the disk. Considering the the total energy contour map (FWHM = 20\arcsec), the minor axis extent of the X-ray emission never exceeds 1.0\,kpc (inclination corrected). The three well resolved point sources in the western halo visible in the total energy band, are not associated with NGC\,3877. Using the ALADIN interactive sky atlas, these objects turn out to be background galaxies. The two X-ray bright point sources close to R.A.\,(J2000)=$11^{\rm h}46^{\rm m}04\fs0$ and Dec.\,(J2000)=$47\degr27\arcmin00\farcs0$ have no optical counterparts and are apparently not associated with NGC\,3877. 
The point source $\sim30\arcsec$ south of the nucleus (best seen in the hard energy band) is the brightest source throughout all bands and can be identified with a star forming region \citep[cf.][]{Ro03b} containing a SNR, emanating from type IIn SN\,1998S \citep{Fa01,Po02}.
As this \ion{H}{ii}-region seems to be embedded in prominent DIG structures, enhanced star formation in the disk plane is likely to occur.   

Therefore, the non-detections of gaseous halos for NGC\,3877 seem to support our hypothesis that multi-phase halos coexist if a certain energy input rate by SNe into the ambient ISM is exceeded. One could argue that the {\small EPIC} images of NGC\,3877 are not deep enough to detect the soft X-ray halo after the GTI-correction has been applied. This, however, can be easily ruled out by considering that NGC\,891 has comparable $S_{60}/S_{100}$ and $L_{\rm FIR}/D^{2}_{25}$ ratios and a significantly larger foreground $N_{\rm H}$, but a positive detection after only 8\,ks of integration time. 
At present, NGC\,3877 is the only galaxy in our sample which neither shows extraplanar radio continuum nor a DIG or a X-ray halo. 

This result poses two important questions: Firstly, do the non-detections indicate that the energy input or equivalently the SFR of NGC\,3877, is too low to drive multi-phase halos? Secondly, if NGC\,3877 should indeed possess a CR-halo, is this component produced first, even at low SFRs? 
For CR halos to evolve, the CR diffusion needs to be high. This, however, would imply that star formation must have been ongoing for some time which on the other hand should have triggered the formation of DIG and X-ray halos.    
An answer to these questions requires to derive the critical threshold SFR above which multi-phase halos form and to consider the evolution and typical formation timescales of each halo component. 

\begin{figure*}
\hspace{-.3cm}
\caption{{\small EPIC} X-ray overlays of NGC\,4634 created from merged pn and MOS images in different energy bands and overlaid onto the H$\alpha$ image presented by \citet{Ro00}. Contours start at $3\sigma$ and are shown on a square-root intensity scale between $2.0\times 10^{-6}\,{\rm cts\ s^{-1}\,pix^{-1}}$ and $2.1\times 10^{-5}\,{\rm cts\ s^{-1}\,pix^{-1}}$. In case of NGC\,4634 a length scale of 1\arcmin\ corresponds to 5.6\,kpc.}
\label{F7} 
\end{figure*}
\subsubsection{NGC\,4631}
This target is another well known Sc edge-on starburst galaxy. Apparently, the strong interaction with two neighboring galaxies \citep[e.g.,][]{Ra94} forced NGC\,4631 to undergo intense star formation in the disk as confirmed by H$\alpha$ imaging data \citep{Ra92,Wa01} and high IRAS flux ratios. As a consequence of these violent processes in the plane, extended non-thermal radio continuum \citep{Co87,Hu88,Go94} and X-ray halos \citep{Vo96,Wa01,St04a} could evolve.

Compared to the $L_{\rm FIR}/D^{2}_{25}$ ratios of other starburst galaxies in our sample, NGC\,4631 turns out to have one of the lowest. This could be explained by assuming that the integrated FIR emission produced by hot stars is distributed over a SF area which is, compared to the nuclear starburst case, substantially larger. This would directly imply that, similar to NGC\,3044, the starburst in NGC\,4631 is not restricted to the nucleus, but affects a large fraction of the disk plane. 

Moreover, this galaxy shows, similar to NGC\,4666 \citep{Da97} and NGC\,5775 \citep{Tu00}, magnetic field vectors in the halo oriented perpendicular to those in the disk \citep{Go94}. This is interpreted as outflowing gas produced by the star forming sites.

The surface brightness of the observed DIG structures \citep{Wa01} seems to be too faint to be traced out to similar distances as observed in NGC\,5775 \citep[e.g.,][]{Tu00,Ra00}. As the reported 16\,kpc detection by \citet{Do95} still needs to be confirmed, we assume a $|z|$-extent of the DIG-halo of about 5.4\,kpc (based on \citet{MK01}, but adjusted to the distance of $D=7.5$\,Mpc). 

Fig.~\ref{F6} presents first XMM-Newton contour images for all energy bands overlaid onto the H$\alpha$ image of \citet{Go96}. The most prominent feature of this galaxy is a huge soft and relatively smooth X-ray halo surrounding the whole disk. The northern maximum $z$-extent has been measured to be 9.1\,kpc, whereas the extension of the southern one is with 5.8\,kpc significantly smaller. These length scales are also consistent with the detected radio continuum halo \citep{Hu88} and earlier X-ray observations \citep{Vo96}.

Contrary to NGC\,3628 (Fig.~\ref{F4}), NGC\,4631 does not show indications of a nuclear starburst as supported by the lack of medium/hard X-ray emission in the innermost 3\,kpc. We find no evidence of AGN-activity in the center of NGC\,4631 \citep{Go99} and confirm previous findings reported by \citet{Vo96} and \citet{St04a}. 
This, together with the coexisting huge radio continuum and X-ray halos, suggests NGC\,4631 to be in a ``spatially extended''-starburst rather than in a ``mild central'', starburst phase \citep{Gow94}.

The point-like source, approximately 1\arcmin\ right to the center, is actually a superposition of two discrete sources \citep[named H7 and H8 in][]{Vo96} and turns out to be the brightest X-ray emitter in NGC\,4631. These objects are most likely SNRs located within a prominent \ion{H}{ii}-region rather than high mass X-ray binaries (see \citet{Vo96} for a detailed discussion). 
Two additional point sources are located in the disk within 2\farcm9 eastward to the center of NGC\,4631. Interestingly, the easternmost object at $\alpha$\,(J2000)\,=\,$12^{\rm h}42^{\rm m}16\fs0$ and $\delta$\,(J2000)\,=\,$32\degr32\arcmin55\farcs0$ \citep[H13 in][]{Vo96} is only visible at energies below 1.0\,keV, whereas H12 (the one to the right of H13) is detectable only above this energy. Time variability could not be established for the former object, whereas for the latter it remains unconfirmed. In order to investigate a possible SNR or X-ray binary nature of these sources, a detailed spectroscopic analysis needs to be carried out.    

In this context it is remarkable that the hard point sources to the east and west of the center seen at energies above 2.0\,keV are both located at positions where the diffuse X-ray contours have their steepest gradient. If the SNR nature of the eastern sources could also be established, additional support would be given to star formation induced multi-phase galaxy halos. 

The extraplanar point source 1\farcm9 above the nucleus of NGC\,4631 is most likely a background source whereas H6 ($\alpha$\,(J2000)\,=\,$12^{\rm h}41^{\rm m}46\fs5$ and $\delta$\,(J2000)\,=\,$32\degr33\arcmin29\farcs0$) is not detectable in our observations, hinting at time variability of this object.
Similar to previous studies carried out with ROSAT \citep{Vo96} or the Chandra X-ray observatory \citep{St04a}, NGC\,4627, the dwarf elliptical companion of NGC\,4631, located at $\alpha$\,(J2000)\,=\,$12^{\rm h}41^{\rm m}59\fs73$ and $\delta$\,(J2000)\,=\,$34\degr34\arcmin24\farcs4$, is also not detectable with XMM-Newton. 
\begin{figure}
\caption{Soft {\small EPIC} X-ray contours overlaid onto a DSS B-band image of NGC\,4634. There is a significant amount of emission produced by stars located in the thick (flaring?) disk and extending radially beyond the H$\alpha$ emitting disk plane (ellipse).}
\label{F8} 
\end{figure}

\subsubsection{NGC\,4634}
NGC\,4634 has been observed in three DIG-related studies \citep{Ro00,Tude00,Ot02} with a clear detection of a gaseous DIG-halo, extending up to 1.2\,kpc above the disk plane. Radio continuum measurements for NGC\,4634 are still scarce and only the integrated flux density (flux, hereafter) at 1.4\,GHz has been measured in the NRAO VLA Sky Survey \citep{Co98}. Unfortunately, high resolution maps are not available, but the NVSS-image\footnote{See http://www.cv.nrao.edu/nvss/postage.shtml} ($\theta$\,=\,45\arcsec\ FWHM resolution), indicates at least the existence of extended emission. In addition, the relatively high $L_{\rm FIR}/D^{2}_{25}$ ratio of about $12.4\times 10^{40}$\,erg s$^{-1}$ kpc$^{-2}$ suggests to consider NGC\,4634 to be a starburst galaxy, too. This, however, can be ruled out, as line ratios obtained at several slit positions are consistent with photoionization by O-stars \citep{Tude00}. As there is only little kinematical line broadening, we expect only weak contribution from shocks. All this points at enhanced star formation and also favors the co-existence of multi-wavelength halos.

The very first X-ray observations of NGC\,4634 are presented in Fig.~\ref{F7}, demonstrating that even non-starburst spiral galaxies can possess very extended soft X-ray halos. The soft energy band shows an undisturbed smooth halo which extends about 4.0\,kpc to the east and 2.6\,kpc to the west of the disk plane. It is remarkable that X-rays at energies above 4.5\,keV are completely lacking.

The point source visible at R.A.\,(J2000)=$12^{\rm h}42^{\rm m}41\fs0$ and Dec.\,(J2000)=$14\degr17\arcmin07\farcs0$ in the medium/hard band (2.0\,--\,4.5\,keV) could be a background galaxy, but remains unidentified due to the lack of multi-wavelength data. An issue is also the nature of the extraplanar object at R.A.\,(J2000)=$12^{\rm h}42^{\rm m}39\fs3$ and Dec.\,(J2000)=$14\degr17\arcmin55\farcs0$ labeled ``Patch\,1'' in \citet{Ro00} as it is not detectable in X-rays. 

As confirmed by the H$\alpha$ image (see lower left panel in Fig.~\ref{F7}), the disk of NGC\,4634 is almost completely surrounded by DIG (for a detailed view, see the H$\alpha$ image presented by \citet{Ro00}). A similar layer-like distribution is observable in X-rays by the smooth and symmetric appearance of the soft X-ray contours. From a morphological point of view, the well-ordered matter distribution of the different gaseous phases in NGC\,4634 poses the most significant difference to NGC\,891, the other non-starburst galaxy with extraplanar gas.
Although the overall extent of the halo is significantly smaller than the one of NGC\,891, NGC\,4634 possesses the most distinct X-ray halo of all non-starburst galaxies detected so far.  

Unfortunately, with the available data the relation between star formation and multi-phase halos can hardly be tested, either because of missing high sensitivity radio continuum data or because of the lack of point sources in the disk whose nature could provide a link to star forming related processes. Except for the presence of an extended DIG layer, the only argument in favor of prominent large scale motions perpendicular to the disk, enhanced star formation, and thus of the outflow model, is provided by slight but significant line broadening in the halo of NGC\,4634 \citep{Tude00}. 

Finally, Fig.~\ref{F8} nicely demonstrates that the disk of NGC\,4634 is much more extended than the H$\alpha$ image originally indicated and that the radial extent of the X-ray halo seems to be restricted to the star forming H$\alpha$-disk.   

\begin{figure*}
\hspace{1.0cm}
\caption{{\small EPIC} X-ray contours for NGC\,4666 (between $2.4\times 10^{-6}\,{\rm cts\ s^{-1}\,pix^{-1}}$ and $5.6\times 10^{-5}\,{\rm cts\ s^{-1}\,pix^{-1}}$, contour smoothness = 3 in ds9), overlaid onto the H$\alpha$ image of NGC\,4666 from \citet{Da97}. 1\arcmin\ corresponds to 6.0\,kpc. The position of the ``X''-shaped DIG structure has been highlighted by white dashed lines, so has been SN1965H in the middle right panel.}
\label{F9} 
\end{figure*}

\subsubsection{NGC\,4666}
NGC\,4666 is a starburst galaxy driving a prominent superwind, which is most likely caused by the interaction with its neighbor NGC\,4668 \citep{Da97}. Huge filamentary X-shaped structures are visible in H$\alpha$ on scales of about 8\,kpc (cf. Fig.~\ref{F9}). Moreover, the DIG-halo is associated with a CR-halo (observed at 1.43\,GHz) of similar size. Alike NGC\,5775, magnetic polarization vectors in the halo are oriented perpendicular to the disk \citep{Da97} which we attribute again to outflowing gas produced by high mass SF.
  
A high $L_{\rm FIR}/D^{2}_{25}$ value of $31.9\times 10^{40}$\,erg s$^{-1}$ kpc$^{-2}$ (the highest among our sample) and an enhanced dust temperature of $T_{\rm d}\,>35$\,K \citep{Yo89}, give undoubtedly evidence of the existence of hot stars in the disk of that galaxy. A soft X-ray halo has been detected with ROSAT \citep{Da98}. Unfortunately, the data turned out to be not sensitive enough to derive spectra and probe the multi-temperature structure towards the halo.

The H$\alpha$ image together with relevant X-ray contour maps are shown in Fig.~\ref{F9}. A rather low $N_{\rm H}$ value of $1.74\times 10^{20}$\,cm$^{-2}$ indicates that this galaxy is only little affected by foreground absorption. As can be seen in the supersoft image, the diffuse X-ray emission appears to be less pronounced than in the soft band. Internal absorption likely causes the apparent constriction of the emission along the south-eastern rim of the disk. 

The diffuse X-ray emission from the disk of NGC\,4666 is strong throughout all energies, even in the hard band and traces the central starburst. Most of the extraplanar emission seems to be confined by the filamentary DIG structures. Although the somewhat low inclination of 80$\degr$ makes it difficult to disentangle emission originating in the disk from that at extraplanar distances, it appears to be reasonable to associate the aligned magnetic field vectors as well as the extraplanar H$\alpha$ and X-ray emitting gas to the outflow cone of NGC\,4666.
     
Another aspect which also supports the relation between star formation and gaseous galaxy halos is the detection of a type-II SN \citep[SN1965H,][see mark in the middle right panel of Fig.~\ref{F9}]{Ba99} very close to the position where the western filamentary DIG structures of the northern and southern halo are anchored to the disk plane. 

\subsubsection{NGC\,5775}
This edge-on Sc spiral galaxy has been studied in the radio continuum \citep[e.g.,][]{Du98,Ir99,Tu00}, in \ion{H}{i} \citep{Ir94}, in the FIR \citep{Le01}, as well as in DIG-related studies \citep{Co00,Ra00,Tu00,Ot02}. However, detailed  X-ray observations are still lacking. 

Alike NGC\,4666, this target has comparably high $S_{60}/S_{100}$ and $L_{\rm FIR}/D_{25}^{2}$ ratios which suggest to consider NGC\,5775 as a starburst galaxy \citep[see also][]{lehe}. 
In case of NGC\,5775 the co-existence of radio continuum and DIG-halos was already established on scales of ~8\,--\,10\,kpc \citep{Tu00}, further strengthening the ``starburst'' nature of this galaxy. Similar to NGC\,4666, magnetic B-vectors, which can be traced up to 7.4\,kpc into the halo, are very well aligned with the prominent ``X''-shaped DIG features \citep{Tu00}.

\begin{figure*}
\hspace{-.1cm}
\caption{Merged {\small EPIC} X-ray contours (between $3.3\times 10^{-6}\,{\rm cts\ s^{-1}\,pix^{-1}}$ and $8.5\times 10^{-4}\,{\rm cts\ s^{-1}\,pix^{-1}}$ and spaced by factors of $\sqrt2$) in different energy bands overlaid onto the H$\alpha$ image of NGC\,5775 presented by \citet{Tu00}. In case of NGC\,5775, a linear length scale of 1\arcmin\ corresponds to 7.8\,kpc. Again, the outer edges of the extended outflow cone are highlighted by white dashed lines and the position of SN1996AE is marked by a black cross in the middle right panel.}
\label{F10} 
\end{figure*}

First high sensitivity X-ray contour maps for NGC\,5775 are overplotted onto the H$\alpha$ image from \citet{Tu00} and shown in Fig.~\ref{F10}. There is obviously a good spatial correspondence between the extended X-ray emission and the H$\alpha$ filaments. From a morphological point of view the DIG in NGC\,5775 is similarly shaped as in NGC\,4666 and it appears reasonable to associate these structures with a limb brightened outflow cone. The similarities to NGC\,4666 also hold with respect to the X-ray emission which is most pronounced where it coincides with extraplanar DIG. Furthermore, a type IIn SN \citep[SN1996AE,][]{Va96} has been detected in the southern part of the disk, again close to where the outflow cone is connected to the disk. This source is also visible on our medium/hard contour map and seems to be one of the dominant X-ray emitters in the disk of NGC\,5775.

The southernmost point source visible at medium and hard X-rays in Fig.~\ref{F10} is coincident with a complex of bright \ion{H}{ii}-regions and a strong radio continuum emitter \citep[feature ``A'' in][]{Le01} supporting again significant star formation activity in the disk. 

Although the above arguments are fully consistent with the outflow scenario, there are two additional observational facts which also suggest that the different extraplanar gas components are indeed produced by violent star forming processes in the disk: firstly, the detection of an expanding \ion{H}{i}-supershell coincident with the southeastern filament \citep{Le01} and secondly, by optical spectroscopy carried out along the vertical walls of the outflow cone \citep{Ra00,Tu00} which found the gas to be not rotating at $|z|$-heights $\ge\,5$\,kpc. 
It remains to be investigated whether it is possible for infalling gas to be initially at the systemic velocity, to spin up as it falls down, and to become entrained by galactic material. 

Moreover, infalling gas is expected to produce shocks, which should lead to significant line broadening and an altered ionization structure. Based on results from a Monte Carlo photoionization simulation, we found the halo of NGC\,5775 to be consistent with pure photoionization by OB stars \citep{Tu05}. 

\begin{figure*}
\caption{UV contour maps obtained with the Optical Monitor (OM) telescope aboard XMM-Newton (UVW2 filter, UVW1 for NGC\,891) are plotted onto the previously mentioned H$\alpha$ and DSS images. Contour levels start at $3\,\sigma$ above the background and are spaced by factors of $\sqrt{2}$. Obviously, the UV flux is a good tracer of the stellar sources in the disk. Wherever this emission is peaked, extended extraplanar DIG and X-ray features are visible, indicating that these structures are closely linked to star forming processes in the disk.}
\label{F11} 
\end{figure*}

\setcounter{figure}{10}
\begin{figure*}
\caption{continued}
\end{figure*}

Evidently, the central part of the galaxy is almost deficient of hard X-rays and only the eastern rim of the disk is X-ray bright at energies between 4.5 and 12.0\,keV. The reason remains unexplained. Despite a high molecular gas content within that region \citep{Le01}, internal absorption appears to be unlikely as this effect should vanish at energies above 2.0\,keV.

Another interesting observation is the detection of soft X-rays at the southern end of the disk where a chain of \ion{H}{ii}-regions protrudes off the disk plane. This is also the place where significant radio continuum at 20\,cm has been detected \citep{Le01}. All this could be indicative of extraplanar star formation, likely triggered by the interaction process with NGC\,5774, rather than by a central burst of star formation as supposed in case of NGC\,55 \citep{Tu03}. 

The prominent point source visible in all energy bands at R.A.\,(J2000)=$14^{\rm h}53^{\rm m}56^{\rm s}$ and Dec.\,(J2000)=$03\degr34\arcmin00\farcs0$ is most likely a background galaxy cluster.

\subsection{OM imagery}
From a morphological point of view we showed in the previous sections that, apart from NGC\,3877, DIG, radio continuum, and X-ray emission are well associated in halos of star forming galaxies. As we want to investigate whether or not stellar feedback processes, such as SNe and stellar winds, are responsible for the creation of galaxy halos, the extraplanar gas should be associated with the main sites of star formation in the disk below. 

In order to further test this connection, supplemental UV-data at 210\,nm have been obtained simultaneously to our EPIC observations, using the OM-telescope aboard XMM-Newton. These UV images are a good tracer of the stellar continuum, and are overlaid onto the H$\alpha$ images which are indicative of SF activity and outflowing ionized gas. All UV-contour maps are presented in Fig.~\ref{F11}.
Unfortunately, there are no UV-data available which allow us to cross check the stellar distribution in the disk of NGC\,3877 with the morphology of the DIG. 

As can be best seen for NGC\,3628, NGC\,4634, and NGC\,5775, the UV continuum radiation is seriously affected by dust absorption in the disk plane. Nevertheless, extraplanar DIG emission is well associated with sites where the UV-flux in the disk is enhanced, e.g., in the northern part of NGC\,891. Morphologically, the extended DIG layer in NGC\,4634 correlates well with the chain of star forming regions in the disk \citep[see][]{Ro00}. In case of NGC\,3628, NGC\,4666, and NGC\,5775, one can interpret the extraplanar DIG as the limb brightened walls of giant outflow cones and thus make the connection with the central UV sources which seem to be the main drivers of the outflow.

In all cases a clear correlation between DIG and UV continuum originating in the disk plane is found. As DIG also correlates with diffuse soft X-ray emitting gas in the halo, it is self-evident to also consider a correlation between HIM and stellar feedback processes, as traced by UV radiation.  

Independent confirmation of a good spatial correlation between H$\alpha$, X-ray, and UV emission has been reported by \citet{Ho05} for the starburst galaxies NGC\,253 and M82, observed with the GALEX satellite. 

\subsection{EPIC pn spectroscopy}
\begin{figure*}
\centering
\caption{EPIC pn spectra of NGC\,891. The extracted source (solid lines) and background regions (dashed lines) are shown in the uppermost plot. In the upper left panel the disk spectrum together with residuals of the fit (data minus folded model) are displayed. A spectrum of the eastern halo is provided in the lower left panel. Panels on the right show similar data, but for the western halo. All details of the fit, the adopted model and predicted parameters are listed in Table~3.}
\label{F12} 
\end{figure*}

\begin{figure*}
\centering
\caption{EPIC pn spectra of NGC\,3628. The extracted regions are shown in the uppermost panel. Although the prominent instrumental Al-K line at 1.5\,keV is still present in the disk spectrum after background subtraction, this feature does not affect the derived flux as it is not fitted by the adopted model. Details of the fit can be found in Table~3.}
\label{F13} 
\end{figure*}

\begin{figure*}
\centering
\caption{EPIC pn spectra of NGC\,4631. Details of the fit are presented in 
Table~3.}
\label{F14} 
\end{figure*}

\begin{figure*}
\centering
\caption{EPIC pn spectra of NGC\,4634. Details of the fit are presented in 
Table~3.}
\label{F15} 
\end{figure*}

\begin{figure*}[!thp]
\centering
\caption{EPIC pn spectra of NGC\,4666. Details of the fit are presented in 
Table~3.}
\label{F16} 
\end{figure*}

\begin{figure*}[!thp]
\centering
\caption{EPIC pn spectra of NGC\,5775. Details of the fit are presented in 
Table~3.}
\label{F17} 
\end{figure*}

\begin{figure*}
\centering
\caption{Representative $3\sigma$ confidence contour plots of the best-fit model for the two thermal plasma components, shown for NGC\,3628.}
\label{F18} 
\end{figure*}

In order to also examine the temperature and other essential parameters of the HIM (e.g., electron densities, gas masses, and radiative cooling times) as a function of distance from the disk, XMM-Newton pn-eventlists have been used to extract spectra at different offset positions from the plane. 

As the background was taken locally from individual exposures rather than from {\small 'CLOSED'} observations (filter wheel in 'closed' position), the prominent Al-K emission line at 1.5\,keV was not always completely removed by background subtraction. 

Due to low count rates at extraplanar distances, spectra of the diffuse emission could not be extracted for NGC\,3044, NGC\,3221, and NGC\,3877; hence only the soft X-ray flux in the disk is given. For NGC\,891, NGC\,3628, NGC\,4631, NGC\,4634, NGC\,4666, and NGC\,5775 (Figs.~\ref{F12}\,--\,\ref{F17}) we obtained spectra at several offset positions from the disk plane, which enabled us to investigate critical parameters of the HIM in unprecedented detail. 

Each of these Figures shows the extracted regions used for background subtraction (dashed rectangles) and for scientific analysis (solid rectangles and ellipses). In the present case ellipses highlight the outer boundaries of the H$\alpha$ disks. The labels assigned to the disk and halo regions refer to the corresponding spectra. Each spectrum contains a string ``bkg'' followed by a number which allows to identify the background used for subtraction.  

Spectra are plotted only for energy channels with a significant number of counts per time and energy interval. For the halo this is usually the range between 0.3\,keV\,--\,2\,keV, whereas for the disk the whole energy range is shown (typically between 0.3\,keV and 12\,keV). The upper limit of 2\,keV for the halo nicely demonstrates that the diffuse X-ray halos are indeed very soft. 

In order to investigate the temperature gradient in the halo we fitted all spectra with {\small XSPEC} using a simple 3-component model, consisting of a photoelectric absorber (assuming Wisconsin cross-sections from \citet{MM83}) to account for the foreground absorption and two thermal Raymond-Smith (RS) plasma models \citep{rs77}. Both RS components are fixed to cosmic metal abundances. The individual $N_{\rm H}$ values have been taken from \citet{DL90} and are not allowed to vary during the $\chi^2$-minimization process. 
Derived temperatures and details of the best-fits, such as the reduced $\chi^2$ (goodness-of-fit), the null hypothesis probability, and the degree of freedom are shown in Table~3. Uncertainties of the fitted temperatures have been calculated on a $2\sigma$ level using the 'error' routine in {\small XSPEC}. 
In addition, Fig.~\ref{F18} displays representative $3\sigma$ confidence contour plots of the plasma temperatures for NGC\,3628, indicating that the achieved solutions are unique and stable. 

Based on spectral fitting, the diffuse X-ray flux of each extracted slice was derived (Table~3, Col. (10)) and can be summed up to yield the integrated diffuse X-ray flux density of the entire galaxy in the energy range between 0.3 and 2.0\,keV. 

It is important to point out that the bumps visible at 0.6\,keV and 0.9\,keV, which can be identified with \ion{O}{vii}, \ion{O}{viii}, and \ion{Fe}{L} emission line complexes, cannot be satisfactorily fitted by a single RS-plasma model. This further strengthens the existence of multi-temperature halos and supports previous findings from self-consistent non-equilibrium ionization (NEI) modeling presented by \citet{Bs03}.    

It should also be stressed that the derived temperatures are no ``real'' plasma temperatures. These quantities depend on fits of high ionization species, such as \ion{Fe}{L} or \ion{O}{vii}, which not necessarily represent the characteristic plasma temperature. 
For the sake of simplicity, we use the term ``gas temperatures'' in this regard.  

Moreover, it has been shown by \citet{BS99} that dynamical models which assume collisional ionization equilibrium (CIE), such as RS or MEKAL-models, are inadequate to simulate the real ionization structure of the hot ionized gas, as the assumption of CIE is usually not valid in environments of strong dynamical evolution. Hence, CIE models can only represent a rough first order approximation. 

As a consequence, significantly better spectral fits and temperatures are expected to be achieved by using self-consistent NEI models, which would also allow to constrain element abundances as well as to derive velocities of the outflowing gas.

Detailed NEI-modeling will be subject of future work and is expected to be essential to distinguish between the infall and outflow scenario.

\begin{table*}
\begin{center}
\caption[]{Best-fit parameters and derived fluxes of the diffuse X-ray emission for all extracted regions.}
\begin{tabular}{cccccccccc}
\hline\hline
\noalign{\smallskip}
Galaxy    & region & $|z|$  & $kT_{\rm soft}$ & $kT_{\rm hard}$  & red. $\chi^{2}$ & dof & nhp   & photons           & Flux\\ 
          &        & [kpc]& [keV]           & [keV]            &                 &     &       & [${\rm s^{-1}\,cm^{-2}}$]  & $[{\rm erg\ s^{-1}\,cm^{-2}}]$ \\
(1)       & (2)    & (3)                    & (4)              & (5)             & (6) & (7)   & (8)      & (9)    & (10)\\ 
\noalign{\smallskip}
\hline
          & E \#1  & 2.80 &0.089\,$\pm$\,0.025 & 0.315\,$\pm$\,0.136 & 0.972  &  7  & 0.450 & 1.47E-04 & 1.82E-13\\ 
NGC\,0891 & disk   & 0.0  &0.160\,$\pm$\,0.036 & 0.651\,$\pm$\,0.115 & 0.738  & 10  & 0.689 & 1.21E-04 & 1.93E-13 (5.46E-13)\\
          & W \#1  & 2.80 &0.096\,$\pm$\,0.021 & 0.277\,$\pm$\,0.180 & 1.185  &  3  & 0.314 & 1.39E-04 & 1.90E-13\\
          & W \#2  & 5.50 &0.085\,$\pm$\,0.028 & 0.287\,$\pm$\,0.135 & 0.295  &  3  & 0.961 & 6.81E-05 & 8.02E-14\\
\hline
NGC\,3044 & disk   & 0.0  &0.105\,$\pm$\,0.039 & 0.821\,$\pm$\,0.123 & 2.932  & 10  & 0.009 & 1.96E-05 & 2.41E-14 (3.25E-14)\\
\hline
NGC\,3221 & disk   & 0.0  &0.142\,$\pm$\,0.057 & 0.806\,$\pm$\,0.138 & 2.177  & 10  & 0.016 & 2.33E-05 & 2.70E-14 (3.65E-14)\\
\hline
          & N \#3  & 8.00 &0.081\,$\pm$\,0.053 & 0.227\,$\pm$\,0.130 & 0.763  &  7  & 0.618 & 3.06E-05 & 3.23E-14\\
          & N \#2  & 5.30 &0.154\,$\pm$\,0.043 & 0.244\,$\pm$\,0.044 & 1.398  & 11  & 0.215 & 7.11E-05 & 7.09E-14\\
          & N \#1  & 2.50 &0.151\,$\pm$\,0.035 & 0.271\,$\pm$\,0.115 & 0.996  & 14  & 0.454 & 1.75E-04 & 1.91E-13\\
NGC\,3628 & disk   & 0.0  &0.148\,$\pm$\,0.022 & 0.692\,$\pm$\,0.121 & 1.909  & 29  &$<$0.01& 1.03E-04 & 1.32E-13 (4.73E-13)\\
          & S \#1  & 2.70 &0.084\,$\pm$\,0.142 & 0.276\,$\pm$\,0.034 & 0.504  & 15  & 0.940 & 1.67E-04 & 1.40E-13\\
          & S \#2  & 5.30 &0.076\,$\pm$\,0.021 & 0.235\,$\pm$\,0.033 & 0.975  & 11  & 0.467 & 7.91E-05 & 6.95E-14\\
\hline
NGC\,3877 & disk   & 0.0  &0.115\,$\pm$\,0.038 & 0.804\,$\pm$\,0.152 & 1.532  &  5  & 0.176 & 7.87E-06 & 1.09E-14 (6.09E-14)\\
\hline
          & N \#5  & 10.6 &0.062\,$\pm$\,0.037 & 0.196\,$\pm$\,0.074 & 0.956  &  9  & 0.472 & 3.56E-05 & 2.84E-14\\ 
          & N \#4  & 9.00 &0.069\,$\pm$\,0.032 & 0.245\,$\pm$\,0.041 & 1.868  &  9  & 0.262 & 6.77E-05 & 5.10E-14\\ 
          & N \#3  & 7.40 &0.096\,$\pm$\,0.038 & 0.211\,$\pm$\,0.075 & 1.438  & 11  & 0.321 & 1.44E-04 & 9.73E-14\\ 
          & N \#2  & 5.80 &0.103\,$\pm$\,0.019 & 0.274\,$\pm$\,0.035 & 1.640  & 27  & 0.128 & 1.94E-04 & 1.73E-13\\
          & N \#1  & 4.20 &0.173\,$\pm$\,0.031 & 0.284\,$\pm$\,0.057 & 1.630  & 25  & 0.082 & 2.17E-04 & 2.25E-13\\
NGC\,4631 & disk   & 0.0  &0.234\,$\pm$\,0.012 & 0.867\,$\pm$\,0.019 & 2.257  & 50  &$<$0.01& 3.10E-04 & 3.77E-13 (6.50E-13)\\
          & S \#1  & 2.90 &0.173\,$\pm$\,0.029 & 0.372\,$\pm$\,0.093 & 1.352  & 28  & 0.101 & 1.15E-05 & 1.18E-13\\
          & S \#2  & 4.50 &0.151\,$\pm$\,0.037 & 0.266\,$\pm$\,0.124 & 1.224  & 15  & 0.244 & 5.72E-05 & 5.90E-14\\
          & S \#3  & 6.10 &0.122\,$\pm$\,0.028 & 0.271\,$\pm$\,0.181 & 0.957  & 14  & 0.495 & 2.99E-05 & 2.53E-14\\
\hline

          & E \#2  & 3.30 &0.093\,$\pm$\,0.012 & 0.191\,$\pm$\,0.089 & 0.876  &  3  & 0.452 & 6.46E-06 & 5.86E-15\\
          & E \#1  & 1.90 &0.108\,$\pm$\,0.032 & 0.267\,$\pm$\,0.113 & 0.929  & 13  & 0.541 & 1.22E-05 & 1.08E-14\\
NGC\,4634 & disk   & 0.0  &0.227\,$\pm$\,0.035 & 0.848\,$\pm$\,0.207 & 0.334  &  8  & 0.953 & 5.11E-06 & 5.76E-15 (6.46E-15)\\
          & W \#1  & 2.00 &0.113\,$\pm$\,0.027 & 0.305\,$\pm$\,0.130 & 0.970  &  7  & 0.451 & 9.94E-06 & 9.73E-15\\
\hline

          & E \#2  & 8.90 &0.092\,$\pm$\,0.024 & 0.228\,$\pm$\,0.063 & 0.512  & 15  & 0.936 & 3.25E-05 & 2.82E-14\\
          & E \#1  & 5.40 &0.150\,$\pm$\,0.043 & 0.373\,$\pm$\,0.074 & 0.702  & 22  & 0.842 & 4.46E-05 & 4.75E-14\\
NGC\,4666 & disk   & 0.0  &0.209\,$\pm$\,0.022 & 0.816\,$\pm$\,0.045 & 1.509  & 36  & 0.026 & 8.83E-05 & 1.10E-13 (2.59E-13)\\
          & W \#1  & 4.90 &0.135\,$\pm$\,0.029 & 0.323\,$\pm$\,0.123 & 0.502  & 27  & 0.985 & 7.07E-05 & 7.78E-14\\
          & W \#2  & 8.60 &0.116\,$\pm$\,0.035 & 0.310\,$\pm$\,0.063 & 0.637  & 19  & 0.881 & 2.85E-05 & 2.68E-14\\
\hline
          & E \#2  & 6.90 &0.067\,$\pm$\,0.031 & 0.251\,$\pm$\,0.101 & 0.025  &  5  & 1.000 & 1.79E-05 & 1.86E-14\\
          & E \#1  & 4.20 &0.085\,$\pm$\,0.022 & 0.280\,$\pm$\,0.053 & 0.272  &  8  & 0.975 & 3.90E-05 & 4.23E-14\\
NGC\,5775 & disk   & 0.0  &0.263\,$\pm$\,0.049 & 0.940\,$\pm$\,0.191 & 0.558  & 19  & 0.937 & 3.22E-05 & 4.40E-14 (1.27E-13)\\
          & W \#1  & 4.40 &0.125\,$\pm$\,0.033 & 0.358\,$\pm$\,0.136 & 0.535  & 10  & 0.866 & 4,62E-05 & 4.99E-14\\
          & W \#2  & 7.05 &0.099\,$\pm$\,0.029 & 0.350\,$\pm$\,0.168 & 0.550  &  8  & 0.820 & 2.39E-05 & 2.81E-14\\
\hline
\end{tabular}
\end{center}
\vspace{-0.2cm}
{\small Notes:\quad Col. (1) Galaxy name. Col. (2): Identifier of the extracted region, according to Figs.~\ref{F12}--\ref{F17}. Col. (3): Average distance above the disk of the extracted region. Col. (4): {\it soft} temperature predicted from the 2-temperature-RS-model. Col. (5) Predicted {\it hard} temperature from the same model. All temperatures are within $2\sigma$ confidence intervals. Col. (6): Reduced $\chi^2$, indicates goodness of fit. Col. (7): Degree of freedom. Col. (8): Null hypothesis probability. Col. (9): Number of photons ${\rm s^{-1}\,cm^{-2}}$ within the extracted region. Col. (10): Corresponding diffuse X-ray flux in the range between 0.3 and 2.0\,keV. Values in parentheses give the total flux (0.3\,--\,12.0\,keV) of the disk.}
\label{tab3}
\end{table*}

\subsubsection{Multi-temperature halos}
The need to apply two plasma components of significantly different temperatures (soft and hard) strongly supports the existence of multi-temperature halos.  
In Fig.~\ref{F19} the derived temperatures are plotted as a function of $z$, the distance above the disk. 

Due to the low inclination of NGC\,4666 it is not trivial to distinguish between diffuse emission originating in the disk and the halo. In order to avoid this confusion, we placed the extraction windows ``E\,\#1'' and  ``W\,\#1'' at offset positions where no disk emission is expected, here beginning at $6\farcs6$ ($\approx$\,850\,pc, inclination corrected) to the east and west of the disk (cf. Fig.~\ref{F16}). As a consequence, temperatures below $|z|\le 4.9$\,kpc are not available. 

Temperatures of the hard component in the disks of our galaxies can be as high as $1.1\times 10^7$\,K, whereas an upper value for the soft component can be placed at about $3\times10^6$\,K. It is important to keep in mind that the disk spectra may still contain contributions from unresolved point sources which could lead to somewhat higher plasma temperatures. 
Gas temperatures in the halo do not drop below $2.2\times10^6$\,K (hard) and $7\times10^5$\,K (soft), respectively. 

Unfortunately, the measured ``soft'' and ``hard'' temperature ranges alone constrain neither the outflow nor the infall model, as they are consistent with those derived from infall modeling \citep{To02} and with NEI-simulations of outflowing gas \citep{Bs03}.
However, the most telling trend in Fig.~\ref{F19} is the decline of both temperature components from the disk towards the halo. 

In the following, we test whether or not these temperature gradients are consistent with changing electron densities and gas masses in the halo and if the outflow can be sustained by comparing dynamical flow times with radiative cooling times.  

\begin{table*}
\begin{center}
\caption{Derived plasma parameters}
\begin{tabular}{cccccccccc}
\hline\hline
\noalign{\smallskip}
NGC    & region & $T_{\rm soft}$ & $T_{\rm hard}$ & $L_{\rm X,soft}$ & $n_{\rm e,soft}$ & $n_{\rm e,hard}$ & $\tau_{\rm c,soft}$ & $\tau_{\rm c,hard}$ & $<\hspace{-0.09cm}M_{\rm gas}\hspace{-0.09cm}>$  \\ 
\noalign{\smallskip}
\ldots &        & \multicolumn{2}{c}{[$10^6$\,K]} & [10$^{40}$\,erg\,s$^{-1}$]& \multicolumn{2}{c}{$\times 10^{-3}/\sqrt{f}$ [cm$^{-3}$]}  &  \multicolumn{2}{c}{$\times 10^{8}$ [yr]} & $\times 10^{8} f$ [M$_{\odot}$] \\
\noalign{\smallskip}
(1)       & (2)    & (3)                      & (4)               & (5)                       & (6)              & (7)           & (8)  & (9) & (10)\\ 
\noalign{\smallskip}
\hline
\noalign{\smallskip}
          & E \#1 & 1.032$\pm$0.290 & 3.654$\pm$1.578 & 0.19 & 1.3$\pm$0.3 & 1.6$\pm$0.5 & 1.0$\pm$0.1 & 3.4$\pm$0.7 & 0.17$\pm$0.09\\ 
     0891 & disk  & 1.856$\pm$0.418 & 7.552$\pm$1.334 & 0.21 (0.59) & 1.1$\pm$0.2 & 1.5$\pm$0.3 & 1.6$\pm$0.2 & 11.2$\pm$1.1 & 0.21$\pm$0.10\\
          & W \#1 & 1.114$\pm$0.244 & 3.213$\pm$2.088 & 0.21 & 1.3$\pm$0.2 & 1.6$\pm$0.4 & 1.0$\pm$0.1 & 3.1$\pm$1.2 & 0.16$\pm$0.07\\
          & W \#2 & 0.986$\pm$0.325 & 3.329$\pm$1.566 & 0.09 & 1.1$\pm$0.2 & 1.3$\pm$0.5 & 1.1$\pm$0.1 & 3.9$\pm$0.7 & 0.08$\pm$0.05\\
\hline                                                                                                        
     3044 & disk  & 1.218$\pm$0.452 & 9.524$\pm$1.427 & 0.09 (0.11) & 0.6$\pm$0.1 & 0.8$\pm$0.1 & 2.5$\pm$0.2 & 27.4$\pm$0.4 & 0.17$\pm$0.02\\
\hline                                                                                                        
     3221 & disk  & 1.647$\pm$0.661 & 9.350$\pm$1.601 & 1.07 (1.45) & 0.6$\pm$0.1 & 0.8$\pm$0.1 & 3.4$\pm$0.2 & 26.9$\pm$0.3 & 0.34$\pm$0.03\\
\hline                                                                                                        
          & N \#3 & 0.940$\pm$0.473 & 2.633$\pm$0.546 & 0.04 & 0.6$\pm$0.1 & 0.7$\pm$0.1 & 1.8$\pm$0.3 & 2.2$\pm$0.3 & 0.06$\pm$0.02\\
          & N \#2 & 1.786$\pm$0.245 & 2.830$\pm$0.473 & 0.08 & 0.7$\pm$0.1 & 0.7$\pm$0.1 & 3.1$\pm$0.1 & 3.8$\pm$0.3 & 0.12$\pm$0.04\\
          & N \#1 & 1.752$\pm$0.230 & 3.144$\pm$0.578 & 0.22 & 1.1$\pm$0.1 & 1.1$\pm$0.1 & 1.8$\pm$0.1 & 2.3$\pm$0.2 & 0.19$\pm$0.05\\
     3628 & disk  & 1.867$\pm$0.154 & 9.721$\pm$0.796 & 0.16 (0.56) & 0.7$\pm$0.1 & 0.7$\pm$0.1 & 3.2$\pm$0.1 &6.0$\pm$0.7 & 0.22$\pm$0.06\\
          & S \#1 & 0.974$\pm$0.237 & 3.202$\pm$0.457 & 0.16 & 1.2$\pm$0.2 & 1.2$\pm$0.2 & 1.0$\pm$0.1 & 1.2$\pm$0.2 & 0.14$\pm$0.04\\
          & S \#2 & 0.882$\pm$0.195 & 2.726$\pm$0.351 & 0.08 & 1.0$\pm$0.2 & 1.0$\pm$0.2 & 1.0$\pm$0.1 & 1.4$\pm$0.1 & 0.08$\pm$0.03\\
\hline                                                                                                        
     3877 & disk  & 1.856$\pm$0.418 & 7.552$\pm$1.334 & 0.02 (0.11) & 0.4$\pm$0.1 & 0.5$\pm$0.1 & 5.7$\pm$0.2 & 34.5$\pm$1.7& 0.07$\pm$0.01\\
\hline                                                                                                        
          & N \#5 & 0.719$\pm$0.429 & 2.274$\pm$0.847 & 0.02 & 0.5$\pm$0.2 & 0.6$\pm$0.3 & 1.6$\pm$0.1 & 4.4$\pm$0.1 & 0.01$\pm$$\le$0.01\\ 
          & N \#4 & 0.800$\pm$0.371 & 2.842$\pm$0.476 & 0.03 & 0.7$\pm$0.2 & 0.8$\pm$0.3 & 1.4$\pm$0.1 & 4.1$\pm$0.1 & 0.01$\pm$$\le$0.01\\ 
          & N \#3 & 1.114$\pm$0.441 & 2.448$\pm$0.870 & 0.07 & 0.9$\pm$0.1 & 1.1$\pm$0.2 & 1.5$\pm$0.1 & 2.8$\pm$0.3 & 0.07$\pm$0.02\\ 
          & N \#2 & 1.195$\pm$0.220 & 3.178$\pm$0.406 & 0.12 & 1.3$\pm$0.2 & 1.5$\pm$0.3 & 1.1$\pm$0.1 & 2.5$\pm$0.1 & 0.10$\pm$0.03\\
          & N \#1 & 2.007$\pm$0.360 & 3.294$\pm$0.661 & 0.15 & 1.5$\pm$0.2 & 1.8$\pm$0.3 & 1.6$\pm$0.1 & 2.3$\pm$0.1 & 0.11$\pm$0.04\\  
     4631 & disk  & 2.714$\pm$0.139 &10.06$\pm$0.220 & 0.25 (0.44) & 0.7$\pm$0.1 & 0.9$\pm$0.2 & 4.9$\pm$0.1 &25.1$\pm$0.1 & 0.44$\pm$0.10\\  
          & S \#1 & 2.007$\pm$0.336 & 4.315$\pm$1.137 & 0.08 & 1.0$\pm$0.1 & 1.2$\pm$0.2 & 2.5$\pm$0.1 & 6.5$\pm$0.4 & 0.09$\pm$0.03\\  
          & S \#2 & 1.752$\pm$0.429 & 3.086$\pm$1.438 & 0.04 & 0.8$\pm$0.1 & 0.9$\pm$0.2 & 2.9$\pm$0.1 & 4.2$\pm$0.5 & 0.06$\pm$0.02\\
          & S \#3 & 1.415$\pm$0.325 & 3.144$\pm$2.100 & 0.02 & 0.6$\pm$0.1 & 0.7$\pm$0.1 & 3.1$\pm$0.1 & 5.7$\pm$1.0 & 0.03$\pm$0.01\\
\hline                                                                                                                                
                                                                                                                                      
          & E \#2 & 1.079$\pm$0.139 & 2.216$\pm$1.032 & 0.03 & 1.3$\pm$0.2 & 1.5$\pm$0.4 & 1.0$\pm$0.1 & 2.3$\pm$0.3 & 0.02$\pm$$\le$0.01\\
          & E \#1 & 1.253$\pm$0.371 & 3.097$\pm$1.311 & 0.05 & 1.8$\pm$0.3 & 2.0$\pm$0.6 & 0.8$\pm$0.1 & 2.3$\pm$0.2 & 0.03$\pm$0.01     \\
     4634 & disk  & 2.633$\pm$0.406 & 9.837$\pm$2.401 & 0.03 (0.03) & 1.3$\pm$0.2 & 1.8$\pm$0.3 & 2.5$\pm$0.1 &12.8$\pm$0.9 & 0.03$\pm$$\le$0.01\\    
          & W \#1 & 1.311$\pm$0.313 & 3.538$\pm$1.508 & 0.04 & 1.4$\pm$0.2 & 1.6$\pm$0.5 & 1.1$\pm$0.1 & 3.4$\pm$0.3 & 0.03$\pm$0.02     \\
\hline                                                                                                        
                                                                                                                                      
          & E \#2 & 1.067$\pm$0.278 & 2.645$\pm$0.731 & 0.14 & 0.7$\pm$0.2 & 0.8$\pm$0.2 & 1.7$\pm$0.1 & 4.8$\pm$0.2 & 0.18$\pm$0.09\\
          & E \#1 & 1.740$\pm$0.499 & 4.327$\pm$0.858 & 0.23 & 1.0$\pm$0.2 & 1.1$\pm$0.3 & 2.2$\pm$0.1 & 6.1$\pm$0.2 & 0.25$\pm$0.10\\
     4666 & disk  & 2.424$\pm$0.255 & 9.466$\pm$0.522 & 0.54 (1.26) & 1.0$\pm$0.2 & 1.0$\pm$0.2 & 2.9$\pm$0.1 &21.0$\pm$0.2 & 0.51$\pm$0.19\\   
          & W \#1 & 1.566$\pm$0.336 & 3.747$\pm$1.485 & 0.38 & 1.2$\pm$0.2 & 1.4$\pm$0.3 & 1.5$\pm$0.1 & 4.1$\pm$0.3 & 0.32$\pm$0.12\\
          & W \#2 & 1.346$\pm$0.406 & 3.596$\pm$0.731 & 0.13 & 0.7$\pm$0.2 & 0.8$\pm$0.2 & 2.3$\pm$0.1 & 6.8$\pm$0.2 & 0.16$\pm$0.09\\
\hline                                                                                                                                
          & E \#2 & 0.777$\pm$0.360 & 2.912$\pm$1.172 & 0.16 & 0.6$\pm$0.2 & 0.7$\pm$0.3 & 1.6$\pm$0.1 & 6.7$\pm$0.4 & 0.28$\pm$0.19\\
          & E \#1 & 0.986$\pm$0.255 & 3.248$\pm$0.615 & 0.36 & 0.9$\pm$0.2 & 1.0$\pm$0.3 & 1.3$\pm$0.1 & 4.9$\pm$0.2 & 0.41$\pm$0.22\\
     5775 & disk  & 3.051$\pm$0.568 & 10.90$\pm$2.216 & 0.37 (1.08) & 2.6$\pm$0.1 & 3.6$\pm$0.2 & 1.4$\pm$0.1 &6.9$\pm$1.0 & 1.65$\pm$0.22\\ 
          & W \#1 & 1.450$\pm$0.383 & 4.153$\pm$1.577 & 0.42 & 1.0$\pm$0.2 & 1.1$\pm$0.3 & 1.8$\pm$0.1 & 5.8$\pm$0.6 & 0.45$\pm$0.18\\
          & W \#2 & 1.148$\pm$0.336 & 4.060$\pm$1.949 & 0.24 & 0.7$\pm$0.2 & 0.8$\pm$0.3 & 1.9$\pm$0.1 & 7.6$\pm$0.6 & 0.34$\pm$0.20\\
\hline                            
\end{tabular}
\end{center}
\vspace{-0.2cm}
{\small Notes:\quad Cols. (1) to (2): See Table~3. Cols. (3) and (4): Temperatures for the soft and hard component, calculated from Cols. (4) and (5) of Table~3, using $T\ ({\rm K})=1.16\times 10^7 kT\ ({\rm keV})$, with $kT$ being the energy of a thermal source. Col. (5): Diffuse soft X-ray luminosities (0.3\,--\,2.0\,keV), based on distances listed in the third column of Table~2. Values in parentheses list total diffuse X-ray luminosities of the disk (0.3\,--\,12.0\,keV). Cols. (6) to (9): Electron densities and radiative cooling times for the soft and hard component, respectively (calculation see text). Col. (10): Averaged gas masses within the extracted volume, based on averaged densities calculated from Cols. (8) and (9).}
\end{table*}


\subsubsection{Electron densities}
Based on the gas temperatures and X-ray fluxes determined above and under the assumption that the detected extraplanar diffuse hot gas belongs to the halo, we can derive other essential parameters, such as electron densities, gas masses, and radiative cooling times.

According to the galactic wind and cooling flow models used by \citet{Nu84}, the electron density can be calculated from the X-ray luminosity and the volume $V$ of the emitting region via: 
\begin{equation}
L_{\rm X}=0.81\times \Lambda(T,Z)\,n^{2}_{\rm e}\,f\,V , 
\end{equation} 
where $\Lambda(T,Z)$ is the cooling function depending on gas temperature and element abundance and $f$ is the poorly constrained volume filling factor of the HIM. The appropriate cooling coefficients have been taken from \citet{Ray76} and are also restricted to cosmic abundances, to be consistent with our adopted RS-models. 
For simplicity reasons, we consider the ionized volume $V$ to be cylindrical.

\begin{figure}
\centering
\caption{Derived gas parameters plotted as a function of $z$. Soft and hard temperatures have been derived by fitting the emission spectra with a photoelectric absorber and two RS plasma models of different temperatures. Electron densities, radiative cooling times, and average gas masses within the ionized volume are based on soft X-ray luminosities and the derived temperatures (see text for details).}
\label{F19} 
\end{figure}

For most of the temperatures, i.e. between $(0.7-3.0)\times 10^{6}$\,K, the cooling function is nearly constant and we adopt a value of $\Lambda(T,Z_{\odot})=1.25\times 10^{-22}$\,erg cm$^{3}$ s$^{-1}$. 

Alike gas temperatures, electron densities actually represent averaged densities, as they have been determined from spatially extended regions (see extraction charts in Figs.~\ref{F12}\,--\,\ref{F17}). All densities are listed in Table~4 for the ``soft'' and ``hard'' components, respectively and plotted as a function of $z$ in Fig.~\ref{F19}. 

However, if one assumes that the derived gas temperatures are produced in the same volume of a hot ionized plasma, two different (electron) density domains within this gas seem to be unphysical and would just be a consequence of the adopted CIE approximation. 
Independently thereof, the constancy of the cooling function causes both density domains to coincide within their uncertainties which implies that only one density component should be considered.  

NEI-models \citep[e.g.,][]{Bs03} overcome this problem, as the temperature, density, and the ionization structure are calculated self consistently.
For consistency reasons, ``soft'' and ``hard'' densities were calculated. 

For all galaxies in our sample electron densities decline (or at least remain constant) with increasing distance to the disk plane (Fig.~\ref{F19}, Table~4). 
    
\subsubsection{Radiative cooling times}
From the above mentioned models, radiative cooling times can be calculated according to:
\begin{equation}
\tau_{\rm c}= 2.11\times \frac{3kT}{2}\ \sqrt{\frac{V f}{L_{\rm X} \Lambda(T,Z)}}\ .  
\end{equation} 
If we substitute the X-ray luminosity by Eq.~(1), Eq.~(2) translates into:
\begin{equation}
\tau_{\rm c}= 3.5 \times \frac{kT}{ \Lambda(T,Z)\ n_{\rm e}}\ .   
\end{equation} 

Assuming that adiabatic cooling for a ``slow'' wind is negligible for $z\le7$\,kpc \citep{BS99}, the most efficient way to cool the gas is achieved by means of radiative recombination. Higher (lower) gas densities should increase (decrease) the efficiency of radiative cooling. This anti-correlation between densities and radiative cooling times is nicely supported by the corresponding plots of Fig.~\ref{F19} (cf. also Table~4). 
In ``fast'' winds (e.g. $\ge100$\,km/s), however, adiabatic cooling is expected to dominate over radiative cooling.

\subsubsection{Gas masses}
Finally, the total gas mass within the ionized volume can be written as \citep{Nu84}:
\begin{equation}
M_{\rm gas}= 1.31\times m_{\rm H}\ \sqrt{\frac{L_{\rm X} V f}{\Lambda(T,Z)}},
\end{equation} 
which can be expressed as a function of electron density:
 \begin{equation}
M_{\rm gas}= 1.18\times m_{\rm H}\ n_{\rm e}\ V f\,.
\end{equation}
As both density components agree very well within their uncertainties, the average was taken to calculate the gas mass within the extracted volume (Table~4 and Fig.~\ref{F19}).  

Obviously, enhanced densities and reduced cooling times are directly indicative of the condensated gas mass.  

The gas mass in every extracted slice drops with increasing distance to the disk plane. In order to evaluate the validiy of either the outflow or the infall model, this finding awaits cross-checking with results from NEI-simulations. It remains further to be investigated if infalling gas from the IGM can get accumulated and compressed due to the density gradient at the halo/IGM interface.

It is clear that the ionization and the dynamics of the gas are tightly coupled. The ionization structure and cooling time scales are sensitive to varying dynamical quantities, such as densities and temperatures. This causes changes in the overall energy budget of the plasma which in turn affects the dynamics of the system again.

\subsubsection{Stability of the outflow}
A comparison between the dynamical gas flow time and the radiative cooling time provides insights into the fate of the diffuse halo gas.

The time required for the gas to reach a typical distance of $z$\,=\,7.0\,kpc above the disk (starburst galaxies), can simply be written as:
$\tau_{\rm f}(z)= \int_{z_0}^{z} dz/u(z)$. Under the assumption that the flow starts at $z_{0}=1$\,kpc, the outflow velocity $u(z)$ in the investigated region behaves approximatively linearly with $z$, and that a typical average velocity for starburst galaxies is of the order of $u_0=270$\,km/s \citep[see][]{Bs03}, we find $\tau_{\rm f}=0.22\times 10^8$\,yrs.
In actively star forming spirals, the flow time required for the gas to reach $z=3.5$\,kpc and 5.5\,kpc (cf. Table~3, Col.~(2)), respectively, \citep[$u_0\approx95$\,km/s, ][]{BS99} ranges between $\tau_{\rm f}=0.24\,-\,0.41\times 10^8$\,yrs.

Typical radiative cooling times range from 0.8\,--\,$7.6\times 10^8$\,yrs and are significantly larger than dynamical flow time scales. Therefore, we expect the halo gas to be sustained for both types of galaxies. For the starburst case it is likely that the outflowing gas also leaves the galaxy; this would not be consistent with the infall model. 
 
In a forthcoming paper, the soft X-ray luminosities are compared to those obtained at other wavelengths. A statistical analysis of the multi-wavelength luminosity correlations will be carried out in order to investigate critical parameters which might be responsible for the creation of coexisting halos, visible at different wavelengths, such as SFRs or galaxy mass/potential.

\section{The global picture}
In our view the evolution of multi-phase halos proceeds as follows: at first there is enhanced FIR emission, indicating the rapid formation of young massive stars enshrouded by dust clouds. Their UV flux is responsible for the FIR emission as the Lyman continuum photons are absorbed by dust grains and re-emitted at infrared wavelengths. During their short lifetime OB stars contribute substantially to the stellar luminosity at optical and UV wavelengths. We expect that most of the planar and extraplanar H$\alpha$ emitting gas is mainly photoionized by these OB-stars \citep[e.g.,][]{Rey84,Tu00,Wo04,Tu05}. 

Typical recombination times of the DIG amount to $\approx 5\times10^7$\,yrs \citep[assuming gas temperatures of 8\,000\,K and gas densities of the order of $10^{-3}$\,cm$^{-3}$ at $|z|=10$\,kpc, see also][]{Tu02}. As recombination of hydrogen in the halo appears to be very unlikely, the ionization of the DIG can easily be maintained by a dilute radiation field escaping the disk plane. 

Collective SNe finally transport gas, energy, and momentum off the disk plane into the halo and trigger the formation of superbubbles or galactic winds visible in H$\alpha$ and at soft X-rays. Halos of ionized hydrogen are a good tracer of the above mentioned disk-halo interaction, whereas extraplanar X-ray emission, traces the HIM generated by X-ray binaries, SNRs, and superwinds. 
In a final step a radio continuum halo should evolve, tracing the synchrotron emission of high energy electrons produced by SNe or SNRs. The creation of a CR-halo should happen after DIG and X-ray halos have formed, because CRs require evolved SNRs or superbubbles to be produced at significant numbers. 

All starburst and non-starburst galaxies of our sample clearly show a correlation between the individual gaseous halo components and support our present view that the coexistence of multi-phase halos is triggered by star formation activity in the disk plane of a galaxy.

In order to explain the non-detections for NGC\,3877, two simple and natural explanations would be to consider that either the energy input into the ISM is too low or that a massive burst of star formation occurred just recently. In case the former holds true, we could establish a first preliminary limit of the SFR below which no multi-phase halos form. If the latter turns out to be correct, we would face the first step in halo formation just before SNe produce the critical pressure required for the hot gas to break out into the halo. 

\section{Summary and conclusions}
In this first part of the study we presented new X-ray imaging and spectroscopic data of a small sample of nine edge-on galaxies obtained with XMM-Newton. The main goals of the present work were (1) to search for X-ray halos in actively star forming and starburst galaxies, (2) to check if they are associated with CR and DIG-halos, (3) to investigate whether or not these halos can be linked to star forming activity in the underlying disk, and (4) to test the hypothesis of a multi-temperature halo.  

In all galaxies, except NGC\,3877, we detected extended soft X-ray halos. Moreover, all galaxies, except NGC\,3877, also possess extended CR and DIG-halos. Supplemental data obtained in the UV with XMM-Newton's OM-telescope at a center wavelength of about 210\,nm clearly indicated that the UV continuum emitted by the young stellar population in the disk is well correlated with extraplanar DIG and the hot X-ray emitting halo gas. Thus, we conclude that the co-existence of multi-phase gaseous halos are a result of star forming activity in the disk plane rather than of infalling gas accreted from the IGM as advocated by \citet{Be00} and \citet{To02}.

We established for the first time that, beside NGC\,891, also another actively star forming spiral galaxy, NGC\,4634, possesses an extended gaseous multi-phase halo. 
High resolution radio continuum data for NGC\,3877 are needed in order to undoubtedly confirm the non-detection of a multi-phase halo. 

It remains to be investigated if non-detections can generally be attributed to low SFRs and whether the SFR is the only sensitive parameter which indicates the existence of a gaseous multi-phase medium in the halo.
The obtained X-ray luminosities will be compared to those measured at different wavelengths in order to search for additional empirical luminosity relations, similar to those found for the FIR and the radio continuum \citep[e.g.,][]{Co92}.

The high sensitivity and large photon collecting area of XMM-Newton allowed us to carry out detailed {\small EPIC}-pn spectroscopy of faint diffuse emission at several offset positions in the halo. In all cases where spectra could be extracted, the temperature drops with increasing distance from the disk. Two RS plasma models of different temperatures were needed in order to fit all spectra statistically well.
Although the adopted CIE-models are not the ideal choice to simulate the HIM, we found convincing evidence that X-ray halos indeed consist of multiple gas components at different temperatures. A more physical model would imply non-equilibrium ionization.

It has further been found that electron gas densities also decrease as a function of distance to the disk plane and that radiative cooling times behave inversely to the density. Alike electron densities, gas masses in every extracted bin also decline towards the outer halo.

The comparison between dynamical flow times and radiative cooling times indicated that the outflow is likely to be sustained in all galaxies. 

Future work will need to involve self-consistent NEI-modeling which will allow us to derive metal abundances and to check the kinematics of the hot ionized halo gas.
\begin{acknowledgements}
This work is supported by Deutsches Zentrum f\"ur Luft-- und Raumfahrt (DLR) through grant 50\,OR\,0102. DB acknowledges support from the Bundesministerium f\"ur Bildung und Forschung (BMBF) by the DLR through grant 50\,OR\,0207 and the Max-Planck-Gesellschaft (MPG).
We thank M. Dahlem for providing the H$\alpha$ image of NGC\,4666 and M. Ehle for his help with SAS-related issues. This research has made use of the Aladin Interactive Sky Atlas and the SIMBAD database operated at CDS, Strasbourg, France. 
\end{acknowledgements}

\end{document}